\documentclass[aps,prx,reprint,letterpaper,showpacs,twocolumns]{revtex4-1}
\usepackage{graphicx}
\usepackage{bm}
\usepackage{amsfonts}
\usepackage[colorlinks,urlcolor=blue,anchorcolor=blue,linkcolor=blue,citecolor=blue,breaklinks=true]{hyperref}

\begin{document}

\title{Quasicrystalline Chern Insulators}
\author{Ai-Lei He$^{1,2}$, Lu-Rong Ding$^{3}$, Yuan Zhou$^{1,4}$, Yi-Fei Wang$^{3}$, and Chang-De Gong$^{3,1,4}$ }
\affiliation{$^1$National Laboratory of Solid State Microstructures and Department of Physics, Nanjing University, Nanjing 210093, China
\\$^2$Institute for Advanced Study, Tsinghua University, Beijing 100084, China
\\$^3$Center for Statistical and Theoretical Condensed Matter Physics, and Department of Physics, Zhejiang Normal University, Jinhua 321004, China
\\$^4$Collaborative Innovation Center of Advanced Microstructures, Nanjing University, Nanjing 210093, China}
\date{\today}

\begin{abstract}
  Chern insulator or quantum anomalous Hall state is a topological state with integer Hall conductivity but in absence of Landau level. It had been well established on various two-dimensional lattices with periodic structure. Here, we report similar Chern insulators can also be realized on the quasicrystal with $5$-fold rotational symmetry. Providing the staggered flux through plaquettes, we propose two types of quasicrystalline Chern insulators. Their topological characterizations are well identified by the robustness of edge states, non-zero real-space Chern number, and  quantized  conductance. We further find the failure of integer conductivity but with quantized Chern number at some special energies. Our study therefore provide a new opportunity to searching topological materials in aperiodic system.

\end{abstract}


\maketitle

{\it Introduction.---} Chern insulator (CI) or Quantum anomalous Hall (QAH) state is a topological state with non-zero Chern number but without symmetry protection. Unlike the conventional integer quantum Hall state with highly degenerated Landau level realized under strong magnetic field at low temperature\cite{Klitzing}, the CIs have no Landau levels and have zero net flux, and therefore have attracted much interest in recent years. Haldane model is the first CI model established on the honeycomb lattice with the staggered flux~\cite{Haldane}, and had been experimentally realized in the ultra-cold atomic system~\cite{Jotzu2014}. So far, CIs had been successively constructed on various periodic lattice by introducing flux, including the checkerboard-lattice model~\cite{CB0,CB1}, the lattice-Dirac model~\cite{Lattice}, the kagom{\'e}-lattice model~\cite{KG0,KG1,KG2,KG3}, the Lieb-lattice model~\cite{Lieb0,Lieb1,Lieb2}, the ruby-lattice model~\cite{Ruby}, the triangular-lattice model~\cite{Triangular}, the star-lattice model~\cite{Star0,Star1,Star2}, the square-octagon-lattice model~\cite{SQOC0,SQOC1}, etc.  In principle, the topology of CIs with translational symmetry can be well characterized by the topological invariant -- the Chern number or TKNN index~\cite{Thouless}, which integrates the Berry curvature over the first Brillouin zone of the lattice with periodic boundary conditions. The chiral edge states emerge on the open boundary of CIs according to the bulk-edge correspondence. These CIs are all constructed on the crystal lattices with the periodic structure and will be referred as \emph{crystalline CIs} below. Recently, the topological states in crystal have been extended to some special geometries, such as the fullerenes~\cite{Moore}, the M\"{o}bius surfaces~\cite{Mobius}, and the singular lattices~\cite{HeAL2,HeAL3}. Some exotic and intriguing features are revealed,  such as the fractional charge near the singularity and many branches of edge excitations~\cite{HeAL2,HeAL3}.  The crystalline topological insulators with some special crystalline symmetry protection are further proposed, for example, the topological crystalline insulators with certain crystal point group symmetry~\cite{TCI}. Those crystalline topological states substantially enrich the families of topological insulators, and open up a new window for electronic devices. Very recently, significant improvement of searching for topological materials efficiently based on the crystal symmetries are developed~\cite{SEA_M1,SEA_M2,SEA_M3}, and thousands of candidates are predicted. Whether similar topological states can be established on the lattice beyond the periodic lattice is surely interesting.

Quasicrystal is a structure with long-range ordered atomic arrangement but without translational symmetry, firstly discovered in the aluminum-manganese alloy with five-fold rotational symmetry in 1984~\cite{Quasicrystal}. In fact, the conception of quasicrystals with five-fold rotational symmetry has a long history far before their experimental discovery. Some designs had been proposed in the early 16th century, such as the D{$\rm{\ddot u}$}rer's pentagonal tiling~\cite{QC1}, Kepler¡¯s tiling~\cite{QC2} and Penrose pentagon pattern~\cite{QC3}. The D{$\rm{\ddot u}$}rer's tiling is one of the simplest patterns, consisting of only diamonds and pentagons. Recently, some topological states in two-dimensional quasicrystals have been proposed, such as the Hofstadter butterfly under the uniform magnetic field~\cite{LCHN2,TopoQC1,TopoQC2}, the weak topological superconductors~\cite{TopoQC3}, the quantum spin Hall states ~\cite{TopoQC4,TopoQC5}, the high order topological states~\cite{TopoQC6,TopoQC61}, and even the topological photonic states~\cite{TopoQC7}. Some real-space topological indices have been developed to characterized the topological nature of these systems without translational symmetry, such as the Kitaev formula~\cite{Kitaev}, C$^{\star}$-algebras~\cite{Calgebra}, the local Chern marker~\cite{LCHN1,LCHN2}, the real-space formulation of weak invariant~\cite{TopoQC3}, and the spin Bott index~\cite{TopoQC4,TopoQC5}, etc.

In this paper, we propose quasicrystalline CIs in D{$\rm{\ddot u}$}rer's tiling with disk geometry. Two kinds of quasicrystalline CIs are constructed by imposing the flux on the plaquettes of either the diamonds (Type-\uppercase\expandafter{\romannumeral1}) or non-adjacent pentagons (Type-\uppercase\expandafter{\romannumeral2}). These quasicrystalline CI states are identified by the non-zero real-space Chern number, and the robust gapless edge states. We further check the topology of quasicrystalline CIs by the integer conductance. Interestingly, a failure of conductance plateau but with non-zero real-space Chern number at special Fermi energy is observed in Type-\uppercase\expandafter{\romannumeral1} quasicrystalline CI, where a core state emerges at the center of quasicrystalline CI lattice. Our study enriches the Chern insulators, and opens up a new window to search the topological materials beyond crystal.

\begin{figure}[!htb]
\includegraphics[scale=0.95]{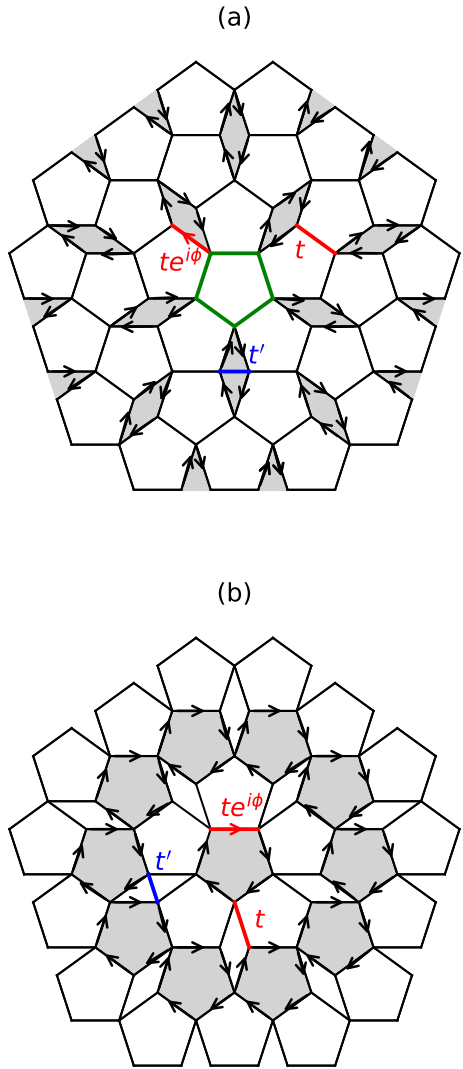}
\caption{(color online). Two types of quasicrystalline CI lattice model realized by imposing the inequivalent staggered flux in D\"{u}rer's tiling with $5$-fold rotational symmetry in the disk geometry. (a) Type-\uppercase\expandafter{\romannumeral1} quasicrystalline CI model. The introduced staggered flux is $-4\phi$ for diamonds (shadow), and $+2\phi$ for pentagons (bright) except for the central one (green without flux). (b) Type-\uppercase\expandafter{\romannumeral2} quasicrystalline CI model. The staggered flux is $-5\phi$ for the non-edge-shared pentagons (shadow), and $+3\phi$ for the remaining pentagons and $+2\phi$ for diamonds (bright), respectively. The adopted gauge is explicitly shown by arrows, which introduce an additional phase factor $\pm\phi$ for part of the nearest neighbor hopping process.}
\label{disk_Lattice}
\end{figure}

{\it Models.---} The $2$-dimentional quasicrystal lattice we adopted is the D{$\rm{\ddot u}$}rer's tiling with $5$-fold rotational symmetry in a disk geometry, consisting of the pentagons and diamonds as shown in Fig.~\ref{disk_Lattice}. To realize the non-trivial topological state, the inequivalent staggered flux is introduced on the polygons of the quasicrystal lattice. There are different ways to construct the quasicrystalline CIs, here we show two typical types. Type-\uppercase\expandafter{\romannumeral1} (Fig.~\ref{disk_Lattice}(a)): the staggered fluxes are imposed on all diamonds (shadow), and all the pentagons (bright) except the central pentagon (green) with specified magnitude, respectively; Type-\uppercase\expandafter{\romannumeral2} (Fig.~\ref{disk_Lattice}(b)): the staggered fluxes are imposed on all the non-edge-shared pentagons (shadow), and the remaining pentagons and all the diamonds (bright) . A special gauge (arrows in Fig.~\ref{disk_Lattice}) is adopted, producing additional phase factor $\pm\phi$ in part of the hopping process between the nearest neighbors. The physical properties are insensitive to the selected gauge. We have checked that the total flux in the whole disk is exactly zero if the regular pentagon shape in Type-\uppercase\expandafter{\romannumeral1} and equal number of cycles of shadow and bright areas in Type-\uppercase\expandafter{\romannumeral2} is considered, slightly differing from the Haldane model with zero flux in a unit cell\cite{Haldane}. It should be reminded that the central pentagon in Type-\uppercase\expandafter{\romannumeral1} quasicrystal model (green pentagon in Fig.~\ref{disk_Lattice} (a)) is special, where the additional phase factor on all bonds is missing.

The real-space Hamiltonian of these two types of quasicrystalline CIs is therefore given by
\begin{equation}
H= -t\sum_{\langle\mathbf{r}\mathbf{r}^{ \prime}\rangle }
a^{\dagger}_{\mathbf{r}^{ \prime}}a_{\mathbf{r}} e^{i\phi_{\mathbf{r}^{ \prime}\mathbf{r}}}
-t^{\prime}\sum_{\lozenge, \langle\mathbf{r}\mathbf{r}^{\prime}\rangle^{\prime}}
a^{\dagger}_{\mathbf{r}^{\prime}}a_{\mathbf{r}},
\label{e.1}
\end{equation}
where $a^{\dagger}_{\mathbf{r}}$ ($a_{\mathbf{r}}$) creates (annihilates) a particle at vertex (site) $\mathbf{r}$, $\langle\mathbf{r}\mathbf{r^{\prime}}\rangle$ runs over all the nearest neighbor sites, and $\lozenge,\langle\mathbf{r}\mathbf{r^{\prime}}\rangle^{\prime}$ denotes the next-nearest-neighbor sites in each diamond. $\phi_{\mathbf{r}^{ \prime}\mathbf{r}}$ is the phase difference between the nearest-neighbor sites as shown in Fig.~\ref{disk_Lattice}. Here, we set the nearest-neighbor hopping $t$ as unit. Since the Hamiltonian is $5$-fold rotational invariant, the angular momentum is a conservation with good quantum number $L$ ($L=0,1,2,3,4$). In analogy to the Haldane model in absence of Semenoff mass\cite{Haldane}, the CIs in quasicrystal proposed here is induced by the staggered flux. Similar CIs had also been realized on the crystalline lattice model, such as the Kagome-lattice\cite{SQOC1} and star-lattice model\cite{Star2}.

\bigskip

\begin{figure*}
  \vspace{-0.1in}
\includegraphics[width=17cm]{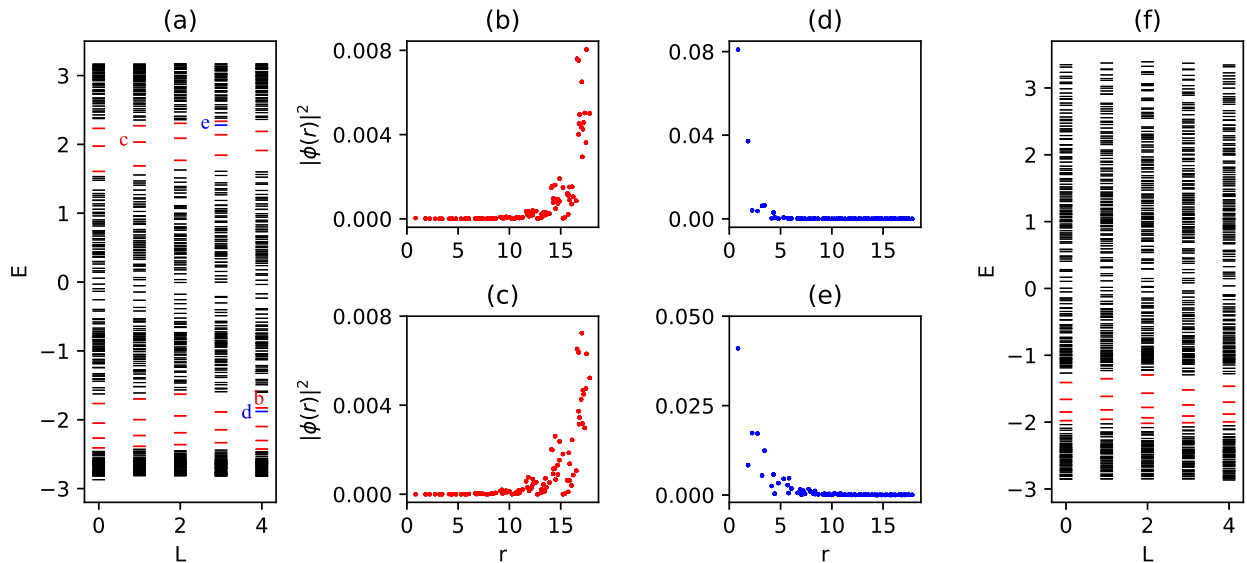}
  \vspace{-0.2in}
\caption{(color online).
Energy spectra of two designed quasicrystalline CIs with $980$ vertexes (or sites). The bulk, edge, and core states, are colored by the black, red, and blue lines, respectively. (a) the spectra of Type-\uppercase\expandafter{\romannumeral1} quasicrystalline CI with $t^{\prime}=-1.3$ and $\phi=\frac{1}{5}\pi$. L is the quantum number of the angular momentum. (b) and (c) are the space distributions of wave function $\vert\psi(r)\vert^{2}$ for edge states highlighted in (a). (d) and (e) are similar $\vert\psi(r)\vert^{2}$ but for the correspondingly core states in (a). (f) The energy spectra for Type-\uppercase\expandafter{\romannumeral2} quasicrystalline CI with $t^{\prime}=0$ and $\phi=-\frac{1}{4}\pi$. No core state exists.
}
\label{single_particle_states}
\end{figure*}

{\it Topological properties.---} To investigate the topology of the above constructed quasicrystal lattice models, we first show the single particle energy spectra in Fig.~\ref{single_particle_states} with $980$ vertexes.  The gapped bulk states with gap about $t$, and the gapless edge states are observed in both type quasicrystal lattice models, in agreement with the the general features in spectra of CI with periodic structure under the open boundary condition~\cite{HeAL2}. The robust edge states are further manifested by the space distribution of wave functions (Fig.~\ref{single_particle_states}) (b) and (c)), which is mainly localized near the boundaries. Interestingly, some additional energies emerge in the bulk gap (blue lines in Fig.~\ref{single_particle_states} (a)) in Type-\uppercase\expandafter{\romannumeral1} quasicrystal model. The space distribution of the wave function analysis indicates that they well locate at the center, i.e., the core states. As mentioned above, the flux phase vanishing in the central pentagon in Type-\uppercase\expandafter{\romannumeral1} quasicrystal system, these core states are indeed the inner "edge states" around the central pentagon. This is quite similar to the core states found in crystalline CI with singular lattices~\cite{HeAL2}. In contrast, no additional core states are observed in Type-\uppercase\expandafter{\romannumeral2} quasicrystal model due to the perfectness of pentagons. Comparison with the energy spectra in the crystalline CIs, we believe the designed quasicrystal systems are quasicrystalline CIs.

\begin{figure}[!htb]
\includegraphics[scale=0.85]{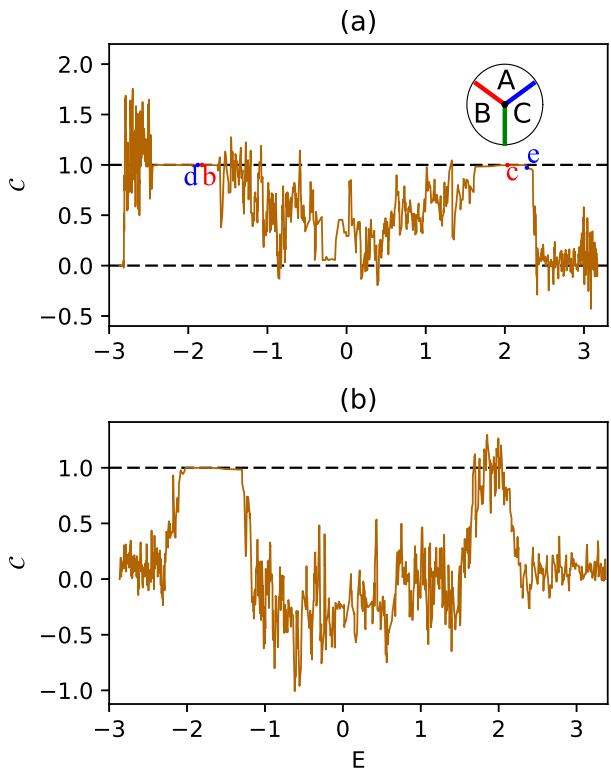}
\caption{(color online). Energy (E) dependence of the number $\mathcal{C}$ for two types of quasicrystalline CIs with fixed $980$ sites. (a) Type-\uppercase\expandafter{\romannumeral1} quasicrystalline CI. There are two real-space Chern number $\mathcal{C}\sim +1$ plateaus. Some selected energies corresponding to the energies highlighted in Fig.~\ref{single_particle_states} (a)) are specified. Insert schematically shows three distinct neighboring regions in the bulk of quasicrystal lattice illustrated in the Kitaev formula as discussed in text. (b) Type-\uppercase\expandafter{\romannumeral2} quasicrystalline CI. There is only one Chern number $\mathcal{C}\sim+1$ plateau. The adopted parameters are same as Fig.~\ref{single_particle_states}.
}

\label{RS_Chern}
\end{figure}

The topological characterization of quasicrystalline CIs can be further identified by the Chern number in the system with time inversion symmetry breaking. Unlike the crystalline CI with periodic lattice, we have to calculate the Chern number in real-space due to the translational symmetry breaking. There are some proposals to compute the real-space Chern number~\cite{Kitaev,Calgebra,LCHN1,LCHN2}. Here, we show the results obtained from the Kitaev formula~\cite{Kitaev} (Appendix~\ref{SM-Kitaev}), and the local Chern number maker (Appendix~\ref{SM-Chernmarker}). The Kitaev formula is expressed as
\begin{equation}\label{Kitaev_formula}
\mathcal{C}=12\pi i \sum_{j\in A} \sum_{k\in B} \sum_{l\in C} (P_{jk}P_{kl}P_{lj}-P_{jl}P_{lk}P_{kj}).
\end{equation}
The disk of quasicrystal lattice is now cut into three distinct neighboring regions arranged in the counterclockwise order shown in insert in Fig.~\ref{RS_Chern}(a). $j, k, l$ denote the vertex (or site) in A, B and C regions, respectively. $\hat{P}$ is the
projection operator defined up to Fermi energy $E_{F}$, i.e., $\hat{P}=\sum_{E_{n}<E_{F}} \vert\phi_{n}\rangle \langle\phi_{n}\vert$, and $P_{jk}=\sum_{E_{n}<E_{F}} \phi_{n}(r_{j})\phi_{n}(r_{k})^{*}$  the matrix elements of $\hat{P}$ with $\phi_n(r_j)=\langle j\vert\phi_{n}\rangle$. The real-space Chern number is independent of the choices of the A, B and C regions~\cite{Sierp}.

According to the Kitaev's proposal in Eq.~(\ref{Kitaev_formula}), the number $\mathcal{C}$ is closely related to the Fermi energy, i.e., $\mathcal{C}\equiv\mathcal{C}({E_F})$. We show the real-space Chern number for two types of quasicrystalline CIs as functions of the Fermi energy with fixed $980$ vertexes in Fig.~\ref{RS_Chern}. Substantial plateaus of quantized real-space Chern number emerge in both types of quasicrystalline CIs when the Fermi energy $E_{F}$ locates in the bulk gap. There are two plateaus in Type-\uppercase\expandafter{\romannumeral1}, and one plateau in Type-\uppercase\expandafter{\romannumeral2} quasicrystalline CI, well consisting with the bulk gap in respective type. To get insight into the Chern number plateaus, we enumerate four specified $E_{F}$ at the fixed energy shown in the spectra (Fig.~\ref{single_particle_states}(a)). The corresponding real-space Chern number is $\mathcal{C}_{b}=0.9991$, $\mathcal{C}_{c}=0.9979$, $\mathcal{C}_{d}=0.9989$, and $\mathcal{C}_{e}=0.9724$, respectively. These Chern numbers are nearly perfect integer except $\mathcal{C}_{e}$, where the core state energy is close to the bulk band. The real-space Chern number is robust against the selected size of quasicrystal lattice in bulk (details see Appendix~\ref{SM-Kitaev}). It should be reminded that the Chern number is a topological invariant, and should be an integer, i.e., $\mathcal{C}=0$ for trivial and $\mathcal{C}\ne 0$ for non-trivial state. The non-integer $\mathcal{C}$ in the metallic state just corresponds to the non-quantized Hall conductivity, and is therefore not a topological invariant.

We also check the real-space Chern number by local Chern number marker scheme, the averaged Chern number in large enough bulk is about $1\pm 0.05$, weakly depending on the selected area (more details see Appendix~\ref{SM-Chernmarker}). Such uncertainty, stemming from the significant inequivalance between the diamonds and pentagons, is directly related to the highly inhomogeneous local Chern number in present quasicrystalline CI models. This is in sharp contrast with the crystalline CIs with equal unit cell, for example, the honeycomb lattice ~\cite{Haldane} and Kagom{\'e} lattice~\cite{HeAL2}. It also differs from the previous suggested quasicrystal model constructed by uniform diamonds, where the local Chern number is homogeneous in bulk~\cite{LCHN2}. However, the average Chern number in large bulk area remains nearly integer, manifesting topology is rather a global property. Similar inhomogeneity is also observed in the densities of many-particle integer filling in the lowest Chern band (Appendix~\ref{SM-MPS}), which can be further understood by the fact that the number of ``bands" cannot be directly associated with the number of atoms in a ``unit cell". Therefore, the inhomogeneity in present quasicrystal models is an intrinsic feature. In this sense, the Kitaev formula is more robust in present cases since it treats the system globally.

{\it Transport property.---}
The integer Hall conductivity is another hallmark of the chiral edge states in CIs, and can be directly identified by the transport measurements. The quantum transport for the non-trivial states have attracted extensive attentions~\cite{Kane,Qi,TopoQC3,TopoQC4,TopoQC5,Kwant} due to its potential applications in electronic devices.

\begin{figure}[!htb]
\includegraphics[scale=0.85]{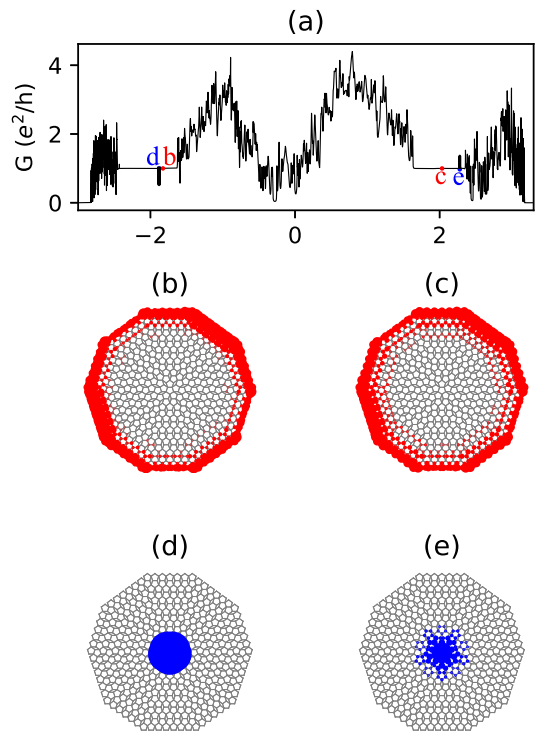}
\caption{(color online). (a) Conductance $G$ (in units of $e^2/h$) with respect to the Fermi energy $E_{F}$ for Type- \uppercase\expandafter{\romannumeral1} quasicrystalline CI. Some specific energies (marked with ``b'', ``c'', ``d'', and ``e'') corresponding to Fig.~\ref{single_particle_states} and \ref{RS_Chern} are highlighted. (b) and (c) The distribution of LDOS $\rho(r,E_{F}$) with $E_{F}$ located in the bulk gap but other than the core state energies. (d) and (e) Similar distribution of LDOS but with $E_{F}$ exactly at the core state energies.}
\label{Transport}
\end{figure}

Here, we perform the quantum transport simulations by using the Kwant. Kwant is a software package for quantum transport and has been widely used to explore transport properties~\cite{Kwant,TopoQC3}. In this toolkit, the conductance $G$ for a disk geometry can be computed between the left lead (L) and the right lead (R) (Fig.~\ref{LDOS} in Appendix~\ref{SM-LDOS}) based on the Landauer-B$\rm{\ddot{u}}$ttiker formula~\cite{Landauer,Buttiker,Datta,transport},
\begin{equation}\label{Landauer_formula}
G=\frac{2e^2}{h}\sum_{m\in{L}}\sum_{n\in{R}}\vert S_{mn}\vert^2,
\end{equation}
where $S_{mn}$ is the scattering matrix and $\vert S_{mn}\vert^2$ denotes the probability that a carrier transmits from the $m$th incoming mode at the left lead to the $n$th outgoing mode at the right lead (more details see Appendix~\ref{SM-LDOS}). However, we can not directly calculate the transverse conductance in disk geometry as that for a multi-terminal rectangular geometry or by using the standard Kubo formula.

The conductance ($G$) as functions of the Fermi energy ($E_{F}$) for type-\uppercase\expandafter{\romannumeral1} quasicrystalline CIs is plotted in Fig.~\ref{Transport}(a).  We add the left and right leads at the edges of the quasicrystal lattice with $980$ vertices (sites). Nearly perfect plateaus with quantized conductance $G=e^2/h$ are observed, in well agreement with the real-space Chern number. The observed conductance quantization in such a two-lead disk geometry can be explained by the formation of a chiral edge channel along the disk edge, and the integer conductance plateau here corresponds to the number of edge channels. Surprisingly, the quantized conductance is significantly broken down at some special $E_{F}$, though the corresponding real-space Chern number remains integer. The corresponding $E_{F}$ is exactly at the energy where the core state located. This failure of quantum conductance is robust against the system size, and cannot be removed by enlarging the quasicrystal lattice (Appendix~\ref{SM-LDOS}), indicating an intrinsic feature for core states.

To get an insight into the breakdown, we show the distribution of the local density of states (LDOS) at specific energies obtained by the Kwant. The LDOS is well located around the center for $E_{F}$ at the core state energy (Fig.~\ref{Transport} (d) and (e)), while it mainly located at the edge for other energies in the bulk gap. In fact, the LDOS discussed here reflects the distribution of moving electrons at the Fermi energy when voltage difference is applied on the leads (Appendix~\ref{SM-LDOS}). The electrons are blockaded by the insulating bulk and edge (white area) in Fig.~\ref{Transport} (d) and (e)), resulting in the failure of quantized conductance. In contrast, they can move dissipationlessly along the edge in Fig.~\ref{Transport} (b) and (c)), preserving the integer conductance.Similar breakdown is also observed in time-reversal symmetry protected topological insulators with magnetic~\cite{Faliue_HC1,Faliue_HC2} or nonmagnetic\cite{Faliue_HC3} impurities due to strong backscattering or antiresonance, but is never reported in the crystalline CI systems.  We believe that the breakdown of conductance quantization should be also found in the CI lattices with singularities~\cite{HeAL2}. In comparison, this failure is not observed in type-\uppercase\expandafter{\romannumeral2} quasicrystalline CI due to the absence of the core states (Appendix~\ref{SM-LDOS}).
\bigskip

{\it Summary and discussion}.---
We propose two types of quasicrystalline CIs by elaborately imposing staggered flux on disk geometry with $5$-fold rotational symmetry in D${\rm{\ddot u}}$rer's pentagonal quasicrystal. Our study therefore enriches the family of CIs, and provides new opportunity to search for topological materials beyond the crystal. The topological properties of these two types of quasicrystalline CIs are well identified by the robust edge states, the non-zero real-space Chern number, and are further checked by the quantized conductance through electronic transport simulations. Interestingly, some core states emerge in the Type-\uppercase\expandafter{\romannumeral1} quasicrystalline CI, where the flux phases vanishing in the central pentagon. These core states cause transmission blockade when the Fermi energy locates at the core states energies, leading to the failure of conductance quantization.

To realize the proposed Chern-insulator states in quasicrystal, a potential way is the ultra-cold atomic system, in which the Haldane model had been experimentally realized~\cite{Jotzu2014}. The quasicrystalline Chern-insulator states may also be simulated by the designed photonic system, in which the topological phase transition~\cite{TP_OPQC} and fractal topological spectrum~\cite{TopoQC7} had been reported. The higher-order topological insulators in quasicrystal was recently proposed to be mapped into an electrical-circuit lattice~\cite{TopoQC61}, providing new feasibility to realize non-trivial topological states. Recently, a dodecagonal quasicrystal was realized in twisted bilayer graphene with $12$-fold rotational symmetry~\cite{Ahn782}, creating new opportunity to find the potential quasicrystalline Chern insulators.
\bigskip

{\it Acknowledgments.---} We thank Jian-Xin Li, Li Sheng, Hong Yao and Qiang-Hua Wang for insightful discussions. This work is supported by the National Natural Science Foundation of China Grant Nos. 11874325 and the Ministry of Science and Technology of China under Grant No. 2016YFA0300401.
\bigskip

\section*{Appendix For ``Quasicrystalline Chern Insulators''}
\setcounter{figure}{0}
\setcounter{equation}{0}
\renewcommand \thefigure{A\arabic{figure}}
\renewcommand \theequation{A\arabic{equation}}
In the main text, we construct two types of quasicrystalline Chern insulator in disk geometry with $5$-fold rotational symmetry in D${\rm{\ddot u}}$rer¡¯s tiling quasicrystal. Here, we show some details about the real-space Chern number calculated by the Kitaev formula~\cite{Kitaev} and the local Chern marker~\cite{LCHN1} for the quasicrystalline CIs, the local density of states through the transport property and some exotic features of particles filling in the Chern band. Differing from the previous CI states in crystalline systems with equivalent unit cell, there are significant inhomogeneities in present quasicrystalline CIs due to the inequivalence of the elemental shapes between the diamonds and pentagons, such as the local Chern markers in the bulk sites, the many-particle densities.

\subsection{Real-space Chern number -- Kitaev formula} \label{SM-Kitaev}
\begin{figure}[!htb]
\includegraphics[scale=0.65]{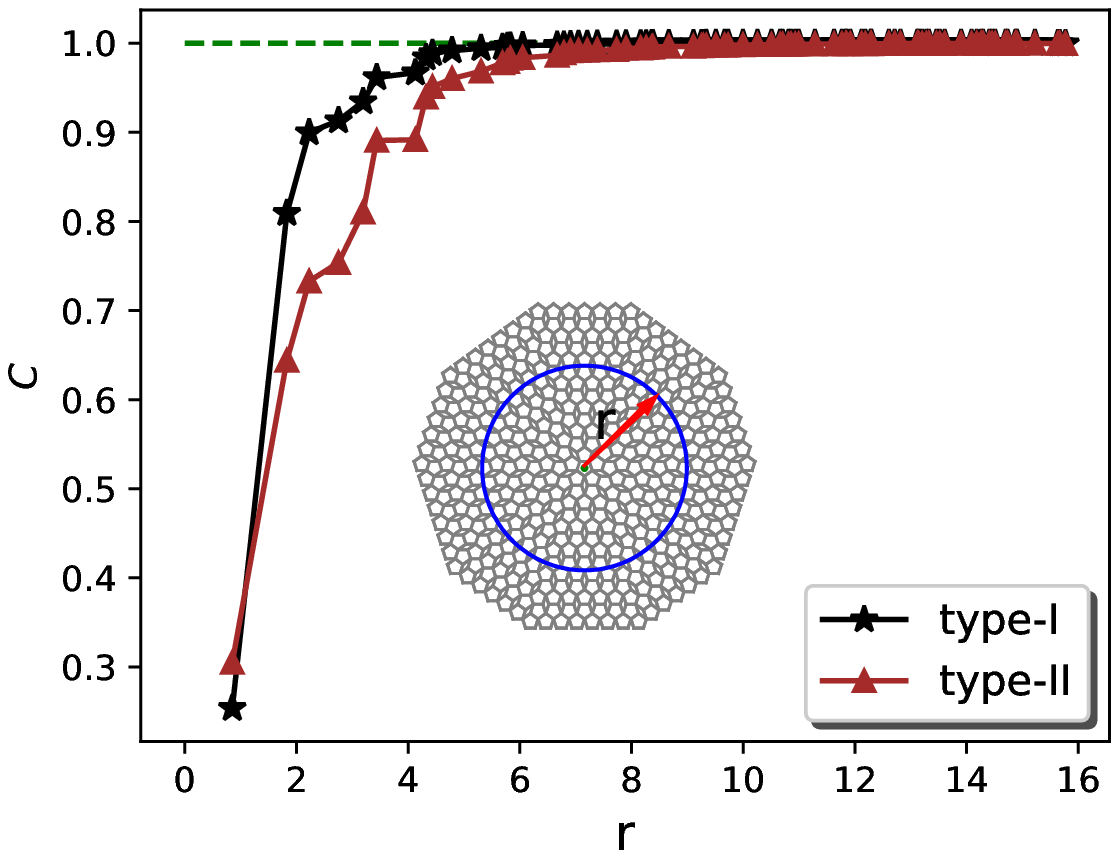}
\caption{(color online). The stability of real-space Chern number $\cal {C}$. We choose the bulk around the center of lattice with varying radius $r$ for two type of quasicrystalline CIs with fixed $980$ sites. The parameters adopted are same as Fig.~\ref{RS_Chern} in main text.}
\label{KVR}
\end{figure}

Chern number, integrating the Berry curvature over the first Brillouin zone, is well defined in the crystalline CIs with periodic structures. In comparison, the real-space Chern number is developed due to the invalidity of Chern number in the translational symmetry breaking system. Here, we show some details about the real-space Chern number using the Kitaev formula~\cite{Kitaev}. The quasicrystal lattice should be cut into three distinct neighboring regions ($A$, $B$ and $C$) arranged in the counterclockwise order. We circle the bulk around the center with radius $r$. The number $\cal {C}$ with varying radius $r$ for two types of quasicrystalline CI is shown in Fig.~\ref{KVR}. $\cal {C}$ is far from $1$ for smaller radius $r<8$, and tends to be a non-zero quantized plateau for larger radius $r>8$. Furthermore, the integer real-space Chern number is robust against the size of quasicrystal lattice if large enough bulk is selected. Since the Kitaev formula treats the bulk as a whole, the corresponding real-space Chern number is indeed a globally invariant number for the bulk, and is in well agreement with the Chern number obtained from the momentum-space.

\subsection{Real-space Chern number -- Local Chern Marker} \label{SM-Chernmarker}
\begin{figure}[!htb]
\includegraphics[scale=0.8]{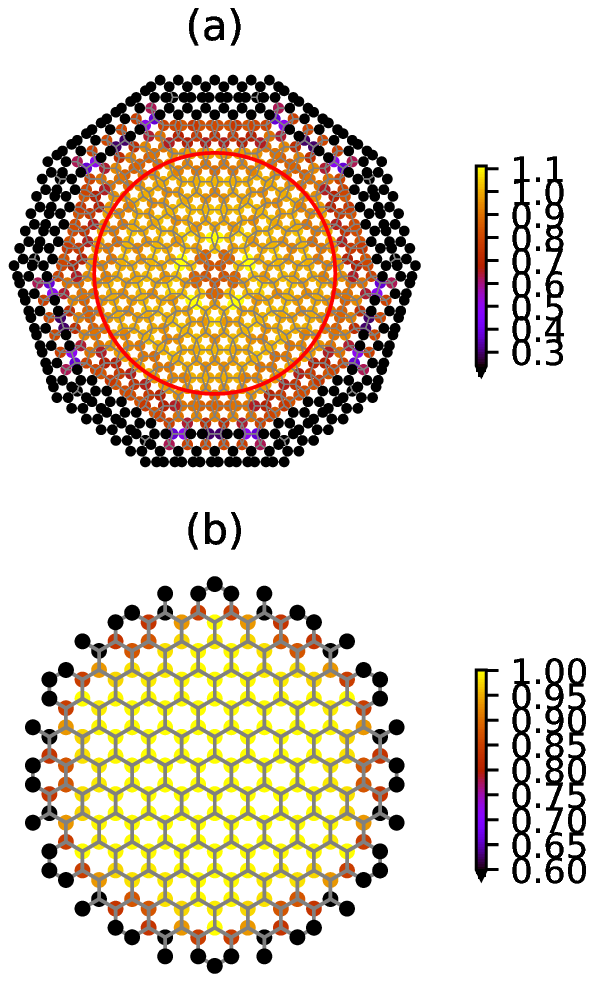}
\caption{(color online). (a). Local Chern marker $\mathfrak{C}(\mathbf{r}_i)$ for Type-\uppercase\expandafter{\romannumeral1} quasicrystalline CI  with fixed $980$ sites. We select a region D with radius $r_D$ in bulk circled by red line. The adopted parameters are same as Fig.~\ref{KVR}. (b). Local Chern marker $\mathfrak{C}(\mathbf{r}_i)$ in the Haldane model.}
\label{LCNM}
\end{figure}
The bulk topological invariant for quasicrystalline CIs can also be calculated by the local Chern marker introduced by Bianco and Resta~\cite{LCHN1}. The details about the general definition of the real-space Chern number $\mathcal{C}$ and the Chern marker in quasicrystal lattice model have been proposed previously~\cite{LCHN1,LCHN2}. Here, we directly apply the expressions of the local Chern number defined at the lattice site $\mathbf{r}_i$~\cite{LCHN2},

\begin{equation}\label{LCHN}
\mathfrak{C}(\mathbf{r}_i)=-4\pi \text{Im} \left [\sum_{\mathbf{r}_j} \langle \mathbf{r}_i \vert \hat{x}_\mathcal{Q} \vert \mathbf{r}_j \rangle \langle \mathbf{r}_j \vert \hat{y}_\mathcal{P} \vert \mathbf{r}_i \rangle \right ]\text{,}
\end{equation}
where
\begin{equation}\label{LCHN_1}
 \langle \mathbf{r}_i \vert \hat{x}_\mathcal{Q} \vert \mathbf{r}_j \rangle=\sum_{\mathbf{r}_k}  \mathcal{Q}(\mathbf{r}_i,\mathbf{r}_k)x_k \mathcal{P}(\mathbf{r}_k,\mathbf{r}_j) ,
\end{equation}
\begin{equation}\label{LCHN_2}
 \langle \mathbf{r}_j \vert \hat{y}_\mathcal{P} \vert \mathbf{r}_i \rangle=\sum_{\mathbf{r}_k}  \mathcal{P}(\mathbf{r}_j,\mathbf{r}_k)y_k \mathcal{Q}(\mathbf{r}_k,\mathbf{r}_i)
\end{equation}
with
\begin{equation}\label{LCHN_3}
 \mathcal{P}(\mathbf{r}_i,\mathbf{r}_j)=\sum_{E_\lambda<E_F} \langle \mathbf{r}_i \vert \psi_\lambda \rangle \langle \psi_\lambda \vert \mathbf{r}_j \rangle\text{,}
\end{equation}
\begin{equation}\label{LCHN_4}
 \mathcal{Q}(\mathbf{r}_i,\mathbf{r}_j)=\sum_{E_\lambda > E_F} \langle \mathbf{r}_i \vert \psi_\lambda \rangle \langle \psi_\lambda \vert \mathbf{r}_j \rangle\ .
\end{equation}
Here $\langle \mathbf{r}_i \vert \psi_\lambda \rangle=\psi_\lambda(\mathbf{r}_i)$ is the real-space wave function at the site $\mathbf{r}_i$ with energy $E_\lambda$, $E_F$ is the Fermi energy. We plot the local Chern number $\mathfrak{C}(\mathbf{r}_i)$ in Fig.~\ref{LCNM} (a) with fixed the Fermi energy $E_F=-1.764$ located in the lower bulk gap in Type-\uppercase\expandafter{\romannumeral1} quasicrystalline CI. The local Chern number $\mathfrak{C}(\mathbf{r}_i)$ is highly inhomogeneous even in the bulk. This inhomogeneity stems from the inequivalence between the elemental shapes of the D${\rm{\ddot u}}$rer¡¯s tiling quasicrystal -- the diamonds and pentagons. Similar inhomogeneity can also be found in Type-\uppercase\expandafter{\romannumeral2} quasicrystalline CI. The present inhomogeneity is in sharp contrast with the previous crystalline CI model with same unit cell, such as the Haldane model, where the local Chern number in bulk is homogeneous and equals to $1$ (Fig.~\ref{LCNM}) (b). It also differs from the previous constructed quasicrystal lattice with uniform diamonds, where the local Chern number is almost $1$~\cite{LCHN2}.

The averaged Chern number $\mathcal{C_D}$ in bulk for the quasicrystalline CI lattice can be defined as
\begin{equation}\label{ACN}
\mathcal{C_D}=\frac{1}{A_D}\sum_{\mathbf{r}_i} \mathfrak{C}(\mathbf{r}_i),
\end{equation}
where $A_{D}=\pi {r_D}^2$ is the area of of the selected region $D$ with radius $r_{D}$ in bulk. Here, we show a typical value of the averaged Chern number in the disk with radius $r_D=10.706$ shown in Fig.~\ref{LCNM}(a). The averaged Chern number $\mathcal{C}(r_D)$ is $0.9996$, nearly perfect integer. However, this averaged Chern number is not stable with $\mathcal{C_D}\approx1\pm0.05$, weakly depending on the selected region. In this sense, the real-space Chern number calculated by the Kitaev formula is more robust than the local Chern marker scheme. The former treats the bulk globally while the latter locally instead.

\subsection{Local Density of States and Failure of quantized Conductance} \label{SM-LDOS}
\begin{figure}[!htb]
\includegraphics[scale=0.8]{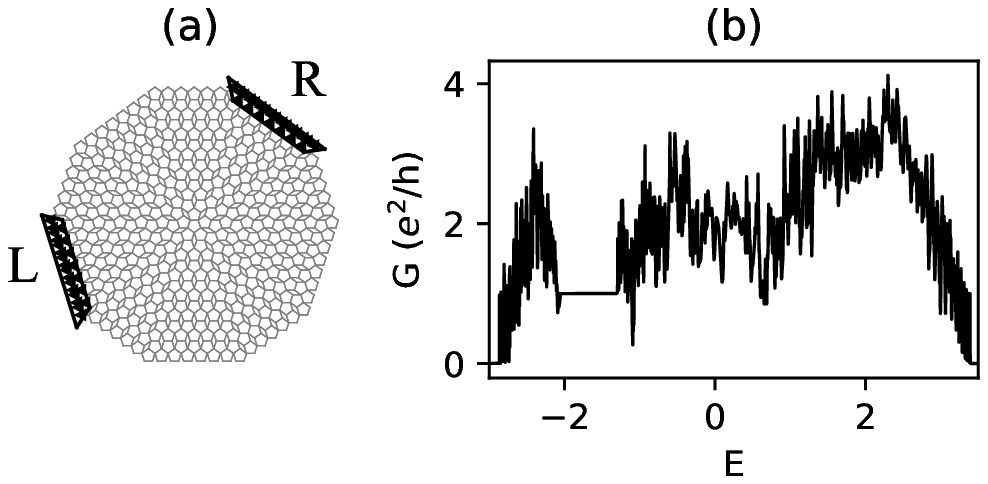}
\caption{(color online). (a). Schematic simulation of the quantum transport using the Kwant. Two leads (L/R) are introduced at the edge of the quasicrystal lattice to calculate the conductance $G$ using the Landauer-B$\rm{\ddot{u}}$ttiker formula. (b). Conductance $G$ (in units of $e^2/h$) with respect to the Fermi energy $E_{F}$ for Type-\uppercase\expandafter{\romannumeral2} quasicrystalline CI.
}
\label{LDOS}
\end{figure}

\begin{figure}[!htb]
\includegraphics[scale=0.8]{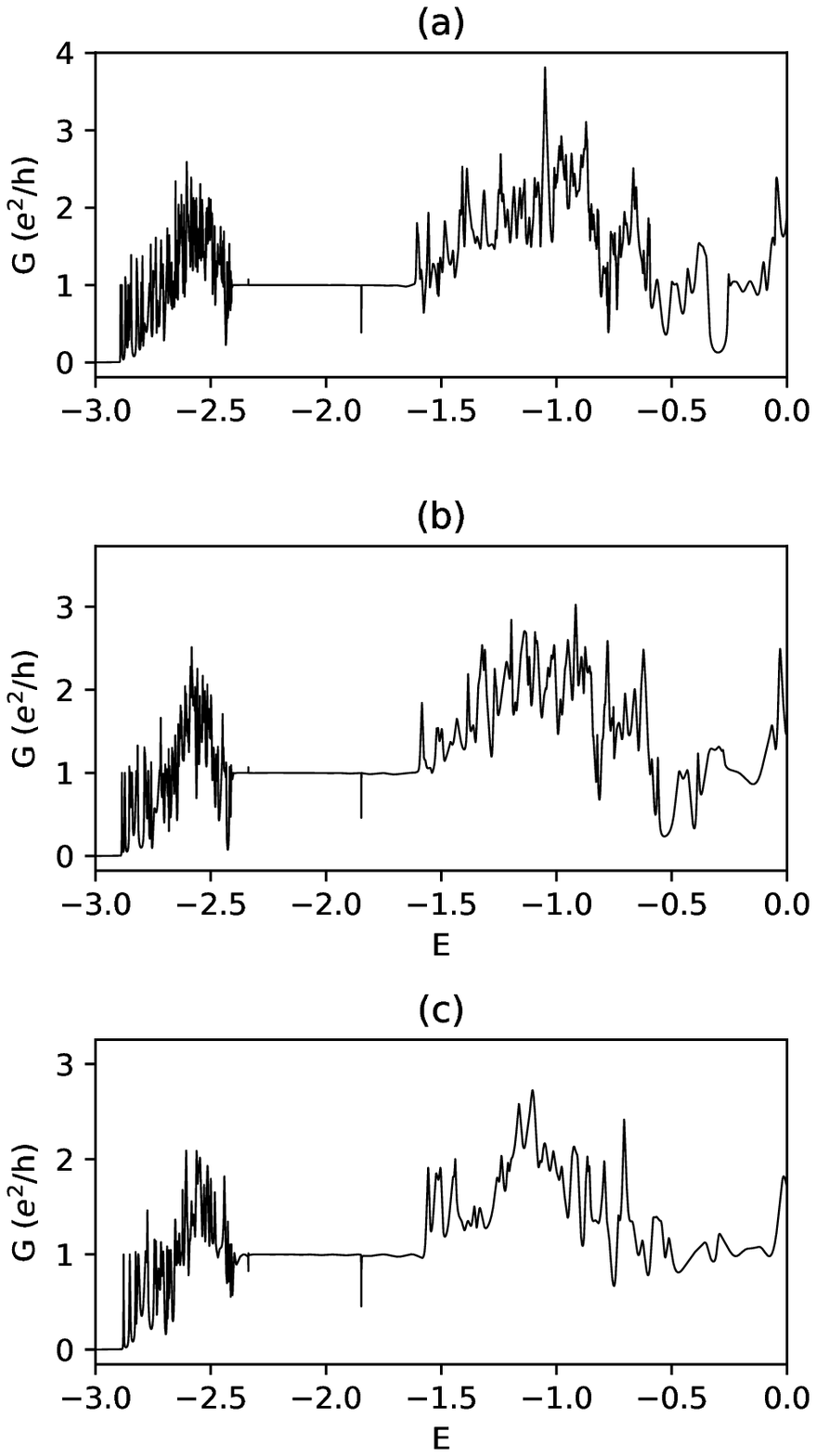}
\caption{(color online). Robustness of the failure of conductance quantization. Here, we choose the first Type of quasicrystalline CI with different sizes, i.e. (a) 720 sites, (b) 500sites and (c) 320 sites.}
\label{Failure_CQ}
\end{figure}

In Fig.~\ref{LDOS} (a), we schematically illustrate the process of the transport simulation by the Kwant. Two leads (L and R) are set at the edge of the quasicrystal lattice. The conductance is therefore calculated by the Landauer-B$\rm{\ddot{u}}$ttiker formula~\cite{Landauer,Buttiker,Datta}
\begin{equation}\label{Landauer_formula}
G=\frac{2e^{2}}{h}\sum_{m\in{L}}\sum_{n\in{R}}\vert S_{mn}\vert^2.
\end{equation}
Here $\vert S_{mn}\vert^{2}$ is the transmitting probability for a carrier from the \emph{m}th incoming mode at the left lead to the \emph{n}th outgoing mode at the right lead with $S_{mn}$ the scattering matrix. This formula as well can be described by the Green¡¯s function method, i.e., $G=\frac{2e^2}{h}T$ with the transmission coefficient $T=$Tr[$\Gamma_L \mathcal{G}^{r}\Gamma_R \mathcal{G}^{a}$]. Here $\Gamma_{L/R}=i[\Sigma^{r}_{L/R}-\Sigma^{a}_{L/R}]$ is the linewidth function defined by the retarded/advanced self-energy ${\Sigma^{r/a}_{L/R}}$. The retard/advance Green function is $\mathcal{G}^{r}=(\mathcal{G}^{a})^{\dagger}=[\mu \mathcal{I}-H_{C}-\Sigma^{r}_{L}-\Sigma^{r}_{R}]^{-1}$ with $H_{C}$ the Hamiltonian matrix of the central scattering region and $\mu$ the chemical potential~\cite{Datta,transport}. The conductance for Type-\uppercase\expandafter{\romannumeral2} quasicrystalline CI is shown in Fig.~\ref{LDOS} (b). Due to the absence of the core state, the conductance plateau well matches the previous plateau in real-space number (Fig.~\ref{RS_Chern} in main text).  On the other hand, we show the breakdown of the quantized conductance with $E_{F}$ at the core state energy though the real-space Chern number remains integer (Fig.~\ref{Transport} in main text). Here, we further show that such failure of quantized conductance is robust against the size of quasicrystal lattices in Fig.~\ref{Failure_CQ}. Therefore, it is an intrinsic property of core states, and cannot be eliminated by enlarging the system size.

\subsection{Many-Particle States} \label{SM-MPS}

\begin{figure}[!htb]
\includegraphics[scale=0.8]{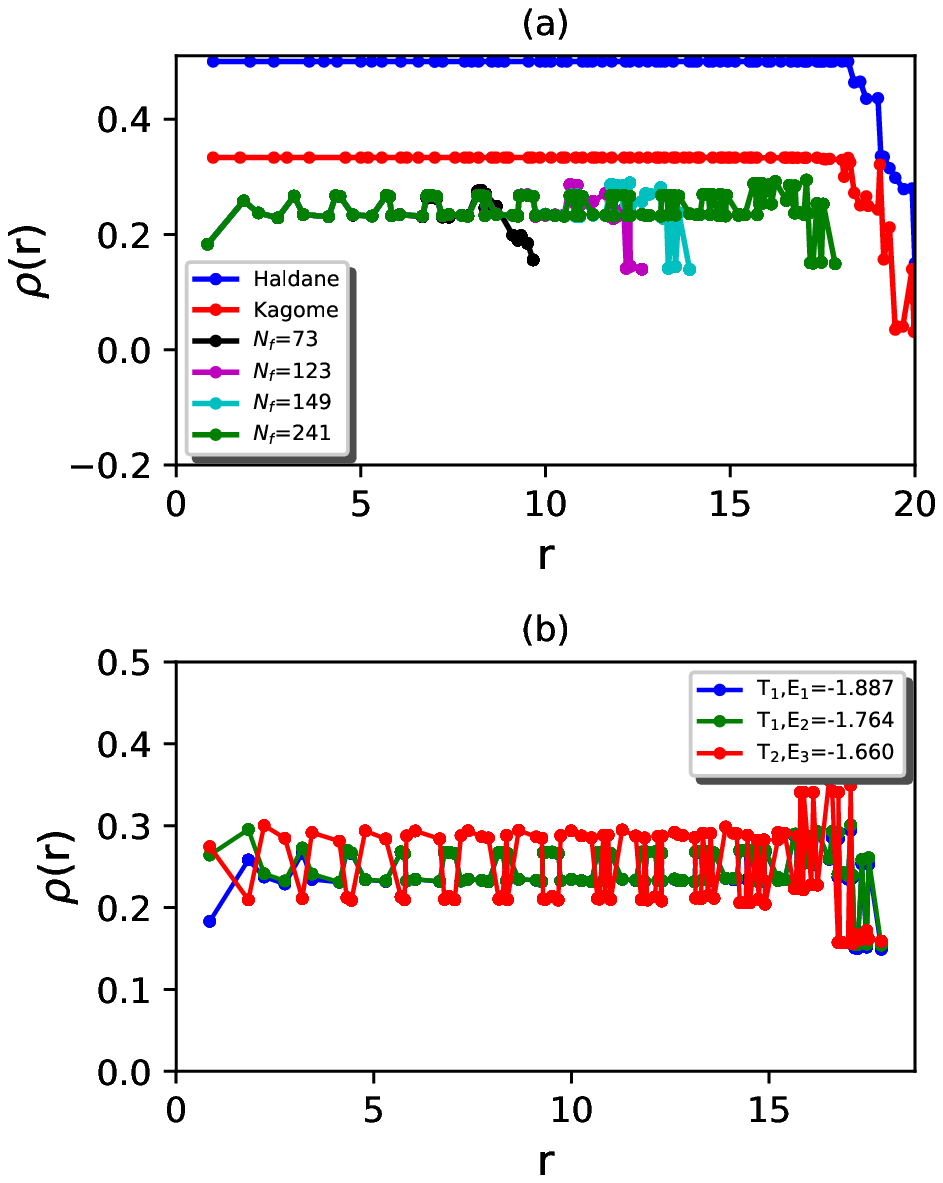}
\caption{(color online). Inhomogeneous many-particle density of quasicrystalline CIs. (a) Many-particle density of quasicrystalline CIs with different sizes of the quasicrystal lattice ($300$, $500$, $600$, and $980$ sites, respectively). For comparison, the many-particle density with particles filling the low energy band ($1/2$ filling, or $1/3$ filling, for Haldane, and Kagom{\'e}, model, respectively) is homogeneous in crystalline CIs (Haldane, Kagom{\'e}, etc). (b) Many-particle density of two types of quasicrystalline CIs with fixed $300$, $500$, $600$ and $980$ sites. For the Type-\uppercase\expandafter{\romannumeral1} quasicrystalline CI, we show two kind of filling with occupying the core states or not.} 
\label{Charge}
\end{figure}

The filling of free spinless fermions in crystalline CIs can be constructed based on the single-particle states and the Pauli principle~\cite{HeAL2}. Here, we show the many-particles states with free spinless fermions filling in the Chern bands in quasicrystal lattice in quasicrystalline CIs. The density of many particles can defined as
\begin{equation}\label{DENS}
\rho(r)=\sum_{E_{\lambda}<E_F}\vert\psi_{\lambda}(r)\vert^2.
\end{equation}
Interestingly, the many-particle densities with near $1/4$ filling are not flat around the center of the disk for both types of quasicrystalline CIs (Fig.~\ref{Charge}), i.e., the many particle densities are inhomogeneous. We also include the many-particle density as function of bulk area with fixed $E_{F}$ in Fig.~\ref{Charge}(b). For Type-\uppercase\expandafter{\romannumeral1} quasicrystalline CI, we choose two Fermi energies below and above the core state energy ($E=-1.8802$), $E_1=-1.887$  and $E_2=-1.764$. For Type-\uppercase\expandafter{\romannumeral2} quasicrystalline CI, we choose the Fermi energy in the bulk gap. All exhibit significant inhomogeneity. In comparison, the many particle density is homogeneous in bulk for the crystalline CIs, such as honeycomb and Kagom{\'e} lattice. As mentioned above, the inhomogeneity originates from the inequivalence between the diamonds and pentagons in present quasicrystalline CIs. The number of ``bands" cannot be directly associated with the number of atoms in a ``unit cell".

\bibliography{QCCIs}

\begin{thebibliography}{58}%
\makeatletter
\providecommand \@ifxundefined [1]{%
 \@ifx{#1\undefined}
}%
\providecommand \@ifnum [1]{%
 \ifnum #1\expandafter \@firstoftwo
 \else \expandafter \@secondoftwo
 \fi
}%
\providecommand \@ifx [1]{%
 \ifx #1\expandafter \@firstoftwo
 \else \expandafter \@secondoftwo
 \fi
}%
\providecommand \natexlab [1]{#1}%
\providecommand \enquote  [1]{``#1''}%
\providecommand \bibnamefont  [1]{#1}%
\providecommand \bibfnamefont [1]{#1}%
\providecommand \citenamefont [1]{#1}%
\providecommand \href@noop [0]{\@secondoftwo}%
\providecommand \href [0]{\begingroup \@sanitize@url \@href}%
\providecommand \@href[1]{\@@startlink{#1}\@@href}%
\providecommand \@@href[1]{\endgroup#1\@@endlink}%
\providecommand \@sanitize@url [0]{\catcode `\\12\catcode `\$12\catcode
  `\&12\catcode `\#12\catcode `\^12\catcode `\_12\catcode `\%12\relax}%
\providecommand \@@startlink[1]{}%
\providecommand \@@endlink[0]{}%
\providecommand \url  [0]{\begingroup\@sanitize@url \@url }%
\providecommand \@url [1]{\endgroup\@href {#1}{\urlprefix }}%
\providecommand \urlprefix  [0]{URL }%
\providecommand \Eprint [0]{\href }%
\providecommand \doibase [0]{https://doi.org/}%
\providecommand \selectlanguage [0]{\@gobble}%
\providecommand \bibinfo  [0]{\@secondoftwo}%
\providecommand \bibfield  [0]{\@secondoftwo}%
\providecommand \translation [1]{[#1]}%
\providecommand \BibitemOpen [0]{}%
\providecommand \bibitemStop [0]{}%
\providecommand \bibitemNoStop [0]{.\EOS\space}%
\providecommand \EOS [0]{\spacefactor3000\relax}%
\providecommand \BibitemShut  [1]{\csname bibitem#1\endcsname}%
\let\auto@bib@innerbib\@empty
\bibitem [{\citenamefont {Klitzing}\ \emph {et~al.}(1980)\citenamefont
  {Klitzing}, \citenamefont {Dorda},\ and\ \citenamefont {Pepper}}]{Klitzing}%
  \BibitemOpen
  \bibfield  {author} {\bibinfo {author} {\bibfnamefont {K.~v.}\ \bibnamefont
  {Klitzing}}, \bibinfo {author} {\bibfnamefont {G.}~\bibnamefont {Dorda}},\
  and\ \bibinfo {author} {\bibfnamefont {M.}~\bibnamefont {Pepper}},\
  }\bibfield  {title} {\bibinfo {title} {New method for high-accuracy
  determination of the fine-structure constant based on quantized hall
  resistance},\ }\href {https://doi.org/10.1103/PhysRevLett.45.494} {\bibfield
  {journal} {\bibinfo  {journal} {Phys. Rev. Lett.}\ }\textbf {\bibinfo
  {volume} {45}},\ \bibinfo {pages} {494} (\bibinfo {year} {1980})}\BibitemShut
  {NoStop}%
\bibitem [{\citenamefont {Haldane}(1988)}]{Haldane}%
  \BibitemOpen
  \bibfield  {author} {\bibinfo {author} {\bibfnamefont {F.~D.~M.}\
  \bibnamefont {Haldane}},\ }\bibfield  {title} {\bibinfo {title} {Model for a
  quantum hall effect without landau levels: Condensed-matter realization of
  the "parity anomaly"},\ }\href {https://doi.org/10.1103/PhysRevLett.61.2015}
  {\bibfield  {journal} {\bibinfo  {journal} {Phys. Rev. Lett.}\ }\textbf
  {\bibinfo {volume} {61}},\ \bibinfo {pages} {2015} (\bibinfo {year}
  {1988})}\BibitemShut {NoStop}%
\bibitem [{\citenamefont {Jotzu}\ \emph {et~al.}(2014)\citenamefont {Jotzu},
  \citenamefont {Messer}, \citenamefont {Desbuquois}, \citenamefont {Lebrat},
  \citenamefont {Uehlinger}, \citenamefont {Greif},\ and\ \citenamefont
  {Esslinger}}]{Jotzu2014}%
  \BibitemOpen
  \bibfield  {author} {\bibinfo {author} {\bibfnamefont {G.}~\bibnamefont
  {Jotzu}}, \bibinfo {author} {\bibfnamefont {M.}~\bibnamefont {Messer}},
  \bibinfo {author} {\bibfnamefont {R.}~\bibnamefont {Desbuquois}}, \bibinfo
  {author} {\bibfnamefont {M.}~\bibnamefont {Lebrat}}, \bibinfo {author}
  {\bibfnamefont {T.}~\bibnamefont {Uehlinger}}, \bibinfo {author}
  {\bibfnamefont {D.}~\bibnamefont {Greif}},\ and\ \bibinfo {author}
  {\bibfnamefont {T.}~\bibnamefont {Esslinger}},\ }\bibfield  {title} {\bibinfo
  {title} {Experimental realization of the topological haldane model with
  ultracold fermions},\ }\href {https://doi.org/10.1038/nature13915} {\bibfield
   {journal} {\bibinfo  {journal} {Nature}\ }\textbf {\bibinfo {volume}
  {515}},\ \bibinfo {pages} {237 EP } (\bibinfo {year} {2014})}\BibitemShut
  {NoStop}%
\bibitem [{\citenamefont {Yakovenko}(1990)}]{CB0}%
  \BibitemOpen
  \bibfield  {author} {\bibinfo {author} {\bibfnamefont {V.~M.}\ \bibnamefont
  {Yakovenko}},\ }\bibfield  {title} {\bibinfo {title} {Chern-simons terms and
  $n$ field in haldane's model for the quantum hall effect without landau
  levels},\ }\href {https://doi.org/10.1103/PhysRevLett.65.251} {\bibfield
  {journal} {\bibinfo  {journal} {Phys. Rev. Lett.}\ }\textbf {\bibinfo
  {volume} {65}},\ \bibinfo {pages} {251} (\bibinfo {year} {1990})}\BibitemShut
  {NoStop}%
\bibitem [{\citenamefont {Sun}\ \emph {et~al.}(2011)\citenamefont {Sun},
  \citenamefont {Gu}, \citenamefont {Katsura},\ and\ \citenamefont
  {Das~Sarma}}]{CB1}%
  \BibitemOpen
  \bibfield  {author} {\bibinfo {author} {\bibfnamefont {K.}~\bibnamefont
  {Sun}}, \bibinfo {author} {\bibfnamefont {Z.}~\bibnamefont {Gu}}, \bibinfo
  {author} {\bibfnamefont {H.}~\bibnamefont {Katsura}},\ and\ \bibinfo {author}
  {\bibfnamefont {S.}~\bibnamefont {Das~Sarma}},\ }\bibfield  {title} {\bibinfo
  {title} {Nearly flatbands with nontrivial topology},\ }\href
  {https://doi.org/10.1103/PhysRevLett.106.236803} {\bibfield  {journal}
  {\bibinfo  {journal} {Phys. Rev. Lett.}\ }\textbf {\bibinfo {volume} {106}},\
  \bibinfo {pages} {236803} (\bibinfo {year} {2011})}\BibitemShut {NoStop}%
\bibitem [{\citenamefont {Qi}\ \emph {et~al.}(2006)\citenamefont {Qi},
  \citenamefont {Wu},\ and\ \citenamefont {Zhang}}]{Lattice}%
  \BibitemOpen
  \bibfield  {author} {\bibinfo {author} {\bibfnamefont {X.-L.}\ \bibnamefont
  {Qi}}, \bibinfo {author} {\bibfnamefont {Y.-S.}\ \bibnamefont {Wu}},\ and\
  \bibinfo {author} {\bibfnamefont {S.-C.}\ \bibnamefont {Zhang}},\ }\bibfield
  {title} {\bibinfo {title} {Topological quantization of the spin hall effect
  in two-dimensional paramagnetic semiconductors},\ }\href
  {https://doi.org/10.1103/PhysRevB.74.085308} {\bibfield  {journal} {\bibinfo
  {journal} {Phys. Rev. B}\ }\textbf {\bibinfo {volume} {74}},\ \bibinfo
  {pages} {085308} (\bibinfo {year} {2006})}\BibitemShut {NoStop}%
\bibitem [{\citenamefont {Ohgushi}\ \emph {et~al.}(2000)\citenamefont
  {Ohgushi}, \citenamefont {Murakami},\ and\ \citenamefont {Nagaosa}}]{KG0}%
  \BibitemOpen
  \bibfield  {author} {\bibinfo {author} {\bibfnamefont {K.}~\bibnamefont
  {Ohgushi}}, \bibinfo {author} {\bibfnamefont {S.}~\bibnamefont {Murakami}},\
  and\ \bibinfo {author} {\bibfnamefont {N.}~\bibnamefont {Nagaosa}},\
  }\bibfield  {title} {\bibinfo {title} {Spin anisotropy and quantum hall
  effect in the kagom\'e lattice: Chiral spin state based on a ferromagnet},\
  }\href {https://doi.org/10.1103/PhysRevB.62.R6065} {\bibfield  {journal}
  {\bibinfo  {journal} {Phys. Rev. B}\ }\textbf {\bibinfo {volume} {62}},\
  \bibinfo {pages} {R6065} (\bibinfo {year} {2000})}\BibitemShut {NoStop}%
\bibitem [{\citenamefont {Guo}\ and\ \citenamefont {Franz}(2009)}]{KG1}%
  \BibitemOpen
  \bibfield  {author} {\bibinfo {author} {\bibfnamefont {H.-M.}\ \bibnamefont
  {Guo}}\ and\ \bibinfo {author} {\bibfnamefont {M.}~\bibnamefont {Franz}},\
  }\bibfield  {title} {\bibinfo {title} {Topological insulator on the kagome
  lattice},\ }\href {https://doi.org/10.1103/PhysRevB.80.113102} {\bibfield
  {journal} {\bibinfo  {journal} {Phys. Rev. B}\ }\textbf {\bibinfo {volume}
  {80}},\ \bibinfo {pages} {113102} (\bibinfo {year} {2009})}\BibitemShut
  {NoStop}%
\bibitem [{\citenamefont {Tang}\ \emph {et~al.}(2011)\citenamefont {Tang},
  \citenamefont {Mei},\ and\ \citenamefont {Wen}}]{KG2}%
  \BibitemOpen
  \bibfield  {author} {\bibinfo {author} {\bibfnamefont {E.}~\bibnamefont
  {Tang}}, \bibinfo {author} {\bibfnamefont {J.-W.}\ \bibnamefont {Mei}},\ and\
  \bibinfo {author} {\bibfnamefont {X.-G.}\ \bibnamefont {Wen}},\ }\bibfield
  {title} {\bibinfo {title} {High-temperature fractional quantum hall states},\
  }\href {https://doi.org/10.1103/PhysRevLett.106.236802} {\bibfield  {journal}
  {\bibinfo  {journal} {Phys. Rev. Lett.}\ }\textbf {\bibinfo {volume} {106}},\
  \bibinfo {pages} {236802} (\bibinfo {year} {2011})}\BibitemShut {NoStop}%
\bibitem [{\citenamefont {Liu}\ \emph {et~al.}(2012)\citenamefont {Liu},
  \citenamefont {Chen}, \citenamefont {Wang},\ and\ \citenamefont
  {Gong}}]{KG3}%
  \BibitemOpen
  \bibfield  {author} {\bibinfo {author} {\bibfnamefont {R.}~\bibnamefont
  {Liu}}, \bibinfo {author} {\bibfnamefont {W.-C.}\ \bibnamefont {Chen}},
  \bibinfo {author} {\bibfnamefont {Y.-F.}\ \bibnamefont {Wang}},\ and\
  \bibinfo {author} {\bibfnamefont {C.-D.}\ \bibnamefont {Gong}},\ }\bibfield
  {title} {\bibinfo {title} {Topological quantum phase transitions and
  topological flat bands on the kagom{\'{e}} lattice},\ }\href
  {https://doi.org/10.1088/0953-8984/24/30/305602} {\bibfield  {journal}
  {\bibinfo  {journal} {Journal of Physics: Condensed Matter}\ }\textbf
  {\bibinfo {volume} {24}},\ \bibinfo {pages} {305602} (\bibinfo {year}
  {2012})}\BibitemShut {NoStop}%
\bibitem [{\citenamefont {Weeks}\ and\ \citenamefont {Franz}(2010)}]{Lieb0}%
  \BibitemOpen
  \bibfield  {author} {\bibinfo {author} {\bibfnamefont {C.}~\bibnamefont
  {Weeks}}\ and\ \bibinfo {author} {\bibfnamefont {M.}~\bibnamefont {Franz}},\
  }\bibfield  {title} {\bibinfo {title} {Topological insulators on the lieb and
  perovskite lattices},\ }\href {https://doi.org/10.1103/PhysRevB.82.085310}
  {\bibfield  {journal} {\bibinfo  {journal} {Phys. Rev. B}\ }\textbf {\bibinfo
  {volume} {82}},\ \bibinfo {pages} {085310} (\bibinfo {year}
  {2010})}\BibitemShut {NoStop}%
\bibitem [{\citenamefont {Goldman}\ \emph {et~al.}(2011)\citenamefont
  {Goldman}, \citenamefont {Urban},\ and\ \citenamefont {Bercioux}}]{Lieb1}%
  \BibitemOpen
  \bibfield  {author} {\bibinfo {author} {\bibfnamefont {N.}~\bibnamefont
  {Goldman}}, \bibinfo {author} {\bibfnamefont {D.~F.}\ \bibnamefont {Urban}},\
  and\ \bibinfo {author} {\bibfnamefont {D.}~\bibnamefont {Bercioux}},\
  }\bibfield  {title} {\bibinfo {title} {Topological phases for fermionic cold
  atoms on the lieb lattice},\ }\href
  {https://doi.org/10.1103/PhysRevA.83.063601} {\bibfield  {journal} {\bibinfo
  {journal} {Phys. Rev. A}\ }\textbf {\bibinfo {volume} {83}},\ \bibinfo
  {pages} {063601} (\bibinfo {year} {2011})}\BibitemShut {NoStop}%
\bibitem [{\citenamefont {Beugeling}\ \emph {et~al.}(2012)\citenamefont
  {Beugeling}, \citenamefont {Everts},\ and\ \citenamefont
  {Morais~Smith}}]{Lieb2}%
  \BibitemOpen
  \bibfield  {author} {\bibinfo {author} {\bibfnamefont {W.}~\bibnamefont
  {Beugeling}}, \bibinfo {author} {\bibfnamefont {J.~C.}\ \bibnamefont
  {Everts}},\ and\ \bibinfo {author} {\bibfnamefont {C.}~\bibnamefont
  {Morais~Smith}},\ }\bibfield  {title} {\bibinfo {title} {Topological phase
  transitions driven by next-nearest-neighbor hopping in two-dimensional
  lattices},\ }\href {https://doi.org/10.1103/PhysRevB.86.195129} {\bibfield
  {journal} {\bibinfo  {journal} {Phys. Rev. B}\ }\textbf {\bibinfo {volume}
  {86}},\ \bibinfo {pages} {195129} (\bibinfo {year} {2012})}\BibitemShut
  {NoStop}%
\bibitem [{\citenamefont {Hu}\ \emph {et~al.}(2011)\citenamefont {Hu},
  \citenamefont {Kargarian},\ and\ \citenamefont {Fiete}}]{Ruby}%
  \BibitemOpen
  \bibfield  {author} {\bibinfo {author} {\bibfnamefont {X.}~\bibnamefont
  {Hu}}, \bibinfo {author} {\bibfnamefont {M.}~\bibnamefont {Kargarian}},\ and\
  \bibinfo {author} {\bibfnamefont {G.~A.}\ \bibnamefont {Fiete}},\ }\bibfield
  {title} {\bibinfo {title} {Topological insulators and fractional quantum hall
  effect on the ruby lattice},\ }\href
  {https://doi.org/10.1103/PhysRevB.84.155116} {\bibfield  {journal} {\bibinfo
  {journal} {Phys. Rev. B}\ }\textbf {\bibinfo {volume} {84}},\ \bibinfo
  {pages} {155116} (\bibinfo {year} {2011})}\BibitemShut {NoStop}%
\bibitem [{\citenamefont {Wang}\ \emph {et~al.}(2012)\citenamefont {Wang},
  \citenamefont {Yao}, \citenamefont {Gong},\ and\ \citenamefont
  {Sheng}}]{Triangular}%
  \BibitemOpen
  \bibfield  {author} {\bibinfo {author} {\bibfnamefont {Y.-F.}\ \bibnamefont
  {Wang}}, \bibinfo {author} {\bibfnamefont {H.}~\bibnamefont {Yao}}, \bibinfo
  {author} {\bibfnamefont {C.-D.}\ \bibnamefont {Gong}},\ and\ \bibinfo
  {author} {\bibfnamefont {D.~N.}\ \bibnamefont {Sheng}},\ }\bibfield  {title}
  {\bibinfo {title} {Fractional quantum hall effect in topological flat bands
  with chern number two},\ }\href {https://doi.org/10.1103/PhysRevB.86.201101}
  {\bibfield  {journal} {\bibinfo  {journal} {Phys. Rev. B}\ }\textbf {\bibinfo
  {volume} {86}},\ \bibinfo {pages} {201101} (\bibinfo {year}
  {2012})}\BibitemShut {NoStop}%
\bibitem [{\citenamefont {Yao}\ and\ \citenamefont {Kivelson}(2007)}]{Star0}%
  \BibitemOpen
  \bibfield  {author} {\bibinfo {author} {\bibfnamefont {H.}~\bibnamefont
  {Yao}}\ and\ \bibinfo {author} {\bibfnamefont {S.~A.}\ \bibnamefont
  {Kivelson}},\ }\bibfield  {title} {\bibinfo {title} {Exact chiral spin liquid
  with non-abelian anyons},\ }\href
  {https://doi.org/10.1103/PhysRevLett.99.247203} {\bibfield  {journal}
  {\bibinfo  {journal} {Phys. Rev. Lett.}\ }\textbf {\bibinfo {volume} {99}},\
  \bibinfo {pages} {247203} (\bibinfo {year} {2007})}\BibitemShut {NoStop}%
\bibitem [{\citenamefont {R\"uegg}\ \emph {et~al.}(2010)\citenamefont
  {R\"uegg}, \citenamefont {Wen},\ and\ \citenamefont {Fiete}}]{Star1}%
  \BibitemOpen
  \bibfield  {author} {\bibinfo {author} {\bibfnamefont {A.}~\bibnamefont
  {R\"uegg}}, \bibinfo {author} {\bibfnamefont {J.}~\bibnamefont {Wen}},\ and\
  \bibinfo {author} {\bibfnamefont {G.~A.}\ \bibnamefont {Fiete}},\ }\bibfield
  {title} {\bibinfo {title} {Topological insulators on the decorated honeycomb
  lattice},\ }\href {https://doi.org/10.1103/PhysRevB.81.205115} {\bibfield
  {journal} {\bibinfo  {journal} {Phys. Rev. B}\ }\textbf {\bibinfo {volume}
  {81}},\ \bibinfo {pages} {205115} (\bibinfo {year} {2010})}\BibitemShut
  {NoStop}%
\bibitem [{\citenamefont {Chen}\ \emph {et~al.}(2012)\citenamefont {Chen},
  \citenamefont {Liu}, \citenamefont {Wang},\ and\ \citenamefont
  {Gong}}]{Star2}%
  \BibitemOpen
  \bibfield  {author} {\bibinfo {author} {\bibfnamefont {W.-C.}\ \bibnamefont
  {Chen}}, \bibinfo {author} {\bibfnamefont {R.}~\bibnamefont {Liu}}, \bibinfo
  {author} {\bibfnamefont {Y.-F.}\ \bibnamefont {Wang}},\ and\ \bibinfo
  {author} {\bibfnamefont {C.-D.}\ \bibnamefont {Gong}},\ }\bibfield  {title}
  {\bibinfo {title} {Topological quantum phase transitions and topological flat
  bands on the star lattice},\ }\href
  {https://doi.org/10.1103/PhysRevB.86.085311} {\bibfield  {journal} {\bibinfo
  {journal} {Phys. Rev. B}\ }\textbf {\bibinfo {volume} {86}},\ \bibinfo
  {pages} {085311} (\bibinfo {year} {2012})}\BibitemShut {NoStop}%
\bibitem [{\citenamefont {Kargarian}\ and\ \citenamefont
  {Fiete}(2010)}]{SQOC0}%
  \BibitemOpen
  \bibfield  {author} {\bibinfo {author} {\bibfnamefont {M.}~\bibnamefont
  {Kargarian}}\ and\ \bibinfo {author} {\bibfnamefont {G.~A.}\ \bibnamefont
  {Fiete}},\ }\bibfield  {title} {\bibinfo {title} {Topological phases and
  phase transitions on the square-octagon lattice},\ }\href
  {https://doi.org/10.1103/PhysRevB.82.085106} {\bibfield  {journal} {\bibinfo
  {journal} {Phys. Rev. B}\ }\textbf {\bibinfo {volume} {82}},\ \bibinfo
  {pages} {085106} (\bibinfo {year} {2010})}\BibitemShut {NoStop}%
\bibitem [{\citenamefont {Liu}\ \emph {et~al.}(2013)\citenamefont {Liu},
  \citenamefont {Chen}, \citenamefont {Wang},\ and\ \citenamefont
  {Gong}}]{SQOC1}%
  \BibitemOpen
  \bibfield  {author} {\bibinfo {author} {\bibfnamefont {X.-P.}\ \bibnamefont
  {Liu}}, \bibinfo {author} {\bibfnamefont {W.-C.}\ \bibnamefont {Chen}},
  \bibinfo {author} {\bibfnamefont {Y.-F.}\ \bibnamefont {Wang}},\ and\
  \bibinfo {author} {\bibfnamefont {C.-D.}\ \bibnamefont {Gong}},\ }\bibfield
  {title} {\bibinfo {title} {Topological quantum phase transitions on the
  kagom{\'{e}} and square{\textendash}octagon lattices},\ }\href
  {https://doi.org/10.1088/0953-8984/25/30/305602} {\bibfield  {journal}
  {\bibinfo  {journal} {Journal of Physics: Condensed Matter}\ }\textbf
  {\bibinfo {volume} {25}},\ \bibinfo {pages} {305602} (\bibinfo {year}
  {2013})}\BibitemShut {NoStop}%
\bibitem [{\citenamefont {Thouless}\ \emph {et~al.}(1982)\citenamefont
  {Thouless}, \citenamefont {Kohmoto}, \citenamefont {Nightingale},\ and\
  \citenamefont {den Nijs}}]{Thouless}%
  \BibitemOpen
  \bibfield  {author} {\bibinfo {author} {\bibfnamefont {D.~J.}\ \bibnamefont
  {Thouless}}, \bibinfo {author} {\bibfnamefont {M.}~\bibnamefont {Kohmoto}},
  \bibinfo {author} {\bibfnamefont {M.~P.}\ \bibnamefont {Nightingale}},\ and\
  \bibinfo {author} {\bibfnamefont {M.}~\bibnamefont {den Nijs}},\ }\bibfield
  {title} {\bibinfo {title} {Quantized hall conductance in a two-dimensional
  periodic potential},\ }\href {https://doi.org/10.1103/PhysRevLett.49.405}
  {\bibfield  {journal} {\bibinfo  {journal} {Phys. Rev. Lett.}\ }\textbf
  {\bibinfo {volume} {49}},\ \bibinfo {pages} {405} (\bibinfo {year}
  {1982})}\BibitemShut {NoStop}%
\bibitem [{\citenamefont {R\"uegg}\ \emph {et~al.}(2013)\citenamefont
  {R\"uegg}, \citenamefont {Coh},\ and\ \citenamefont {Moore}}]{Moore}%
  \BibitemOpen
  \bibfield  {author} {\bibinfo {author} {\bibfnamefont {A.}~\bibnamefont
  {R\"uegg}}, \bibinfo {author} {\bibfnamefont {S.}~\bibnamefont {Coh}},\ and\
  \bibinfo {author} {\bibfnamefont {J.~E.}\ \bibnamefont {Moore}},\ }\bibfield
  {title} {\bibinfo {title} {Corner states of topological fullerenes},\ }\href
  {https://doi.org/10.1103/PhysRevB.88.155127} {\bibfield  {journal} {\bibinfo
  {journal} {Phys. Rev. B}\ }\textbf {\bibinfo {volume} {88}},\ \bibinfo
  {pages} {155127} (\bibinfo {year} {2013})}\BibitemShut {NoStop}%
\bibitem [{\citenamefont {Beugeling}\ \emph {et~al.}(2014)\citenamefont
  {Beugeling}, \citenamefont {Quelle},\ and\ \citenamefont
  {Morais~Smith}}]{Mobius}%
  \BibitemOpen
  \bibfield  {author} {\bibinfo {author} {\bibfnamefont {W.}~\bibnamefont
  {Beugeling}}, \bibinfo {author} {\bibfnamefont {A.}~\bibnamefont {Quelle}},\
  and\ \bibinfo {author} {\bibfnamefont {C.}~\bibnamefont {Morais~Smith}},\
  }\bibfield  {title} {\bibinfo {title} {Nontrivial topological states on a
  m\"obius band},\ }\href {https://doi.org/10.1103/PhysRevB.89.235112}
  {\bibfield  {journal} {\bibinfo  {journal} {Phys. Rev. B}\ }\textbf {\bibinfo
  {volume} {89}},\ \bibinfo {pages} {235112} (\bibinfo {year}
  {2014})}\BibitemShut {NoStop}%
\bibitem [{\citenamefont {He}\ \emph {et~al.}(2018)\citenamefont {He},
  \citenamefont {Luo}, \citenamefont {Wang},\ and\ \citenamefont
  {Gong}}]{HeAL2}%
  \BibitemOpen
  \bibfield  {author} {\bibinfo {author} {\bibfnamefont {A.-L.}\ \bibnamefont
  {He}}, \bibinfo {author} {\bibfnamefont {W.-W.}\ \bibnamefont {Luo}},
  \bibinfo {author} {\bibfnamefont {Y.-F.}\ \bibnamefont {Wang}},\ and\
  \bibinfo {author} {\bibfnamefont {C.-D.}\ \bibnamefont {Gong}},\ }\bibfield
  {title} {\bibinfo {title} {Chern insulators in singular geometries},\ }\href
  {https://doi.org/10.1103/PhysRevB.97.045126} {\bibfield  {journal} {\bibinfo
  {journal} {Phys. Rev. B}\ }\textbf {\bibinfo {volume} {97}},\ \bibinfo
  {pages} {045126} (\bibinfo {year} {2018})}\BibitemShut {NoStop}%
\bibitem [{\citenamefont {He}\ \emph {et~al.}(2019)\citenamefont {He},
  \citenamefont {Luo}, \citenamefont {Wang},\ and\ \citenamefont
  {Gong}}]{HeAL3}%
  \BibitemOpen
  \bibfield  {author} {\bibinfo {author} {\bibfnamefont {A.-L.}\ \bibnamefont
  {He}}, \bibinfo {author} {\bibfnamefont {W.-W.}\ \bibnamefont {Luo}},
  \bibinfo {author} {\bibfnamefont {Y.-F.}\ \bibnamefont {Wang}},\ and\
  \bibinfo {author} {\bibfnamefont {C.-D.}\ \bibnamefont {Gong}},\ }\bibfield
  {title} {\bibinfo {title} {Fractional chern insulators in singular
  geometries},\ }\href {https://doi.org/10.1103/PhysRevB.99.165105} {\bibfield
  {journal} {\bibinfo  {journal} {Phys. Rev. B}\ }\textbf {\bibinfo {volume}
  {99}},\ \bibinfo {pages} {165105} (\bibinfo {year} {2019})}\BibitemShut
  {NoStop}%
\bibitem [{\citenamefont {Fu}(2011)}]{TCI}%
  \BibitemOpen
  \bibfield  {author} {\bibinfo {author} {\bibfnamefont {L.}~\bibnamefont
  {Fu}},\ }\bibfield  {title} {\bibinfo {title} {Topological crystalline
  insulators},\ }\href {https://doi.org/10.1103/PhysRevLett.106.106802}
  {\bibfield  {journal} {\bibinfo  {journal} {Phys. Rev. Lett.}\ }\textbf
  {\bibinfo {volume} {106}},\ \bibinfo {pages} {106802} (\bibinfo {year}
  {2011})}\BibitemShut {NoStop}%
\bibitem [{\citenamefont {Zhang}\ \emph {et~al.}(2019)\citenamefont {Zhang},
  \citenamefont {Jiang}, \citenamefont {Song}, \citenamefont {Huang},
  \citenamefont {He}, \citenamefont {Fang}, \citenamefont {Weng},\ and\
  \citenamefont {Fang}}]{SEA_M1}%
  \BibitemOpen
  \bibfield  {author} {\bibinfo {author} {\bibfnamefont {T.}~\bibnamefont
  {Zhang}}, \bibinfo {author} {\bibfnamefont {Y.}~\bibnamefont {Jiang}},
  \bibinfo {author} {\bibfnamefont {Z.}~\bibnamefont {Song}}, \bibinfo {author}
  {\bibfnamefont {H.}~\bibnamefont {Huang}}, \bibinfo {author} {\bibfnamefont
  {Y.}~\bibnamefont {He}}, \bibinfo {author} {\bibfnamefont {Z.}~\bibnamefont
  {Fang}}, \bibinfo {author} {\bibfnamefont {H.}~\bibnamefont {Weng}},\ and\
  \bibinfo {author} {\bibfnamefont {C.}~\bibnamefont {Fang}},\ }\bibfield
  {title} {\bibinfo {title} {Catalogue of topological electronic materials},\
  }\href {https://doi.org/10.1038/s41586-019-0944-6} {\bibfield  {journal}
  {\bibinfo  {journal} {Nature}\ }\textbf {\bibinfo {volume} {566}},\ \bibinfo
  {pages} {475} (\bibinfo {year} {2019})}\BibitemShut {NoStop}%
\bibitem [{\citenamefont {Vergniory}\ \emph {et~al.}(2019)\citenamefont
  {Vergniory}, \citenamefont {Elcoro}, \citenamefont {Felser}, \citenamefont
  {Regnault}, \citenamefont {Bernevig},\ and\ \citenamefont {Wang}}]{SEA_M2}%
  \BibitemOpen
  \bibfield  {author} {\bibinfo {author} {\bibfnamefont {M.~G.}\ \bibnamefont
  {Vergniory}}, \bibinfo {author} {\bibfnamefont {L.}~\bibnamefont {Elcoro}},
  \bibinfo {author} {\bibfnamefont {C.}~\bibnamefont {Felser}}, \bibinfo
  {author} {\bibfnamefont {N.}~\bibnamefont {Regnault}}, \bibinfo {author}
  {\bibfnamefont {B.~A.}\ \bibnamefont {Bernevig}},\ and\ \bibinfo {author}
  {\bibfnamefont {Z.}~\bibnamefont {Wang}},\ }\bibfield  {title} {\bibinfo
  {title} {A complete catalogue of high-quality topological materials},\ }\href
  {https://doi.org/10.1038/s41586-019-0954-4} {\bibfield  {journal} {\bibinfo
  {journal} {Nature}\ }\textbf {\bibinfo {volume} {566}},\ \bibinfo {pages}
  {480} (\bibinfo {year} {2019})}\BibitemShut {NoStop}%
\bibitem [{\citenamefont {Tang}\ \emph {et~al.}(2019)\citenamefont {Tang},
  \citenamefont {Po}, \citenamefont {Vishwanath},\ and\ \citenamefont
  {Wan}}]{SEA_M3}%
  \BibitemOpen
  \bibfield  {author} {\bibinfo {author} {\bibfnamefont {F.}~\bibnamefont
  {Tang}}, \bibinfo {author} {\bibfnamefont {H.~C.}\ \bibnamefont {Po}},
  \bibinfo {author} {\bibfnamefont {A.}~\bibnamefont {Vishwanath}},\ and\
  \bibinfo {author} {\bibfnamefont {X.}~\bibnamefont {Wan}},\ }\bibfield
  {title} {\bibinfo {title} {Comprehensive search for topological materials
  using symmetry indicators},\ }\href
  {https://doi.org/10.1038/s41586-019-0937-5} {\bibfield  {journal} {\bibinfo
  {journal} {Nature}\ }\textbf {\bibinfo {volume} {566}},\ \bibinfo {pages}
  {486} (\bibinfo {year} {2019})}\BibitemShut {NoStop}%
\bibitem [{\citenamefont {Shechtman}\ \emph {et~al.}(1984)\citenamefont
  {Shechtman}, \citenamefont {Blech}, \citenamefont {Gratias},\ and\
  \citenamefont {Cahn}}]{Quasicrystal}%
  \BibitemOpen
  \bibfield  {author} {\bibinfo {author} {\bibfnamefont {D.}~\bibnamefont
  {Shechtman}}, \bibinfo {author} {\bibfnamefont {I.}~\bibnamefont {Blech}},
  \bibinfo {author} {\bibfnamefont {D.}~\bibnamefont {Gratias}},\ and\ \bibinfo
  {author} {\bibfnamefont {J.~W.}\ \bibnamefont {Cahn}},\ }\bibfield  {title}
  {\bibinfo {title} {Metallic phase with long-range orientational order and no
  translational symmetry},\ }\href
  {https://doi.org/10.1103/PhysRevLett.53.1951} {\bibfield  {journal} {\bibinfo
   {journal} {Phys. Rev. Lett.}\ }\textbf {\bibinfo {volume} {53}},\ \bibinfo
  {pages} {1951} (\bibinfo {year} {1984})}\BibitemShut {NoStop}%
\bibitem [{\citenamefont {Durer}(1977)}]{QC1}%
  \BibitemOpen
  \bibfield  {author} {\bibinfo {author} {\bibfnamefont {A.}~\bibnamefont
  {Durer}},\ }\href@noop {} {\emph {\bibinfo {title} {(1525) A Manual of
  Measurement of Lines, Areas and Solids by Means of Compass and Ruler}}}\
  (\bibinfo  {publisher} {Facsimile Edition (Abaris Books, New York)},\
  \bibinfo {year} {(1977)})\ \bibinfo {note} {translated with commentary by W.
  L. Strauss}\BibitemShut {NoStop}%
\bibitem [{\citenamefont {Kepler}(1969)}]{QC2}%
  \BibitemOpen
  \bibfield  {author} {\bibinfo {author} {\bibfnamefont {J.}~\bibnamefont
  {Kepler}},\ }\href@noop {} {\emph {\bibinfo {title} {(1619) Harmonices
  Mundi}}}\ (\bibinfo  {publisher} {Facsimile edition (Forni Editore, Bologna,
  Italy)},\ \bibinfo {year} {(1969)})\BibitemShut {NoStop}%
\bibitem [{\citenamefont {Penrose}(1974)}]{QC3}%
  \BibitemOpen
  \bibfield  {author} {\bibinfo {author} {\bibfnamefont {R.}~\bibnamefont
  {Penrose}},\ }\bibfield  {title} {\bibinfo {title} {The role of aesthetics in
  pure and applird mathematical research},\ }\href@noop {} {\bibfield
  {journal} {\bibinfo  {journal} {Bull. Inst. Math. Appl.}\ }\textbf {\bibinfo
  {volume} {10}},\ \bibinfo {pages} {266} (\bibinfo {year} {1974})}\BibitemShut
  {NoStop}%
\bibitem [{\citenamefont {Tran}\ \emph {et~al.}(2015)\citenamefont {Tran},
  \citenamefont {Dauphin}, \citenamefont {Goldman},\ and\ \citenamefont
  {Gaspard}}]{LCHN2}%
  \BibitemOpen
  \bibfield  {author} {\bibinfo {author} {\bibfnamefont {D.-T.}\ \bibnamefont
  {Tran}}, \bibinfo {author} {\bibfnamefont {A.}~\bibnamefont {Dauphin}},
  \bibinfo {author} {\bibfnamefont {N.}~\bibnamefont {Goldman}},\ and\ \bibinfo
  {author} {\bibfnamefont {P.}~\bibnamefont {Gaspard}},\ }\bibfield  {title}
  {\bibinfo {title} {Topological hofstadter insulators in a two-dimensional
  quasicrystal},\ }\href {https://doi.org/10.1103/PhysRevB.91.085125}
  {\bibfield  {journal} {\bibinfo  {journal} {Phys. Rev. B}\ }\textbf {\bibinfo
  {volume} {91}},\ \bibinfo {pages} {085125} (\bibinfo {year}
  {2015})}\BibitemShut {NoStop}%
\bibitem [{\citenamefont {Fuchs}\ and\ \citenamefont {Vidal}(2016)}]{TopoQC1}%
  \BibitemOpen
  \bibfield  {author} {\bibinfo {author} {\bibfnamefont {J.-N.}\ \bibnamefont
  {Fuchs}}\ and\ \bibinfo {author} {\bibfnamefont {J.}~\bibnamefont {Vidal}},\
  }\bibfield  {title} {\bibinfo {title} {Hofstadter butterfly of a
  quasicrystal},\ }\href {https://doi.org/10.1103/PhysRevB.94.205437}
  {\bibfield  {journal} {\bibinfo  {journal} {Phys. Rev. B}\ }\textbf {\bibinfo
  {volume} {94}},\ \bibinfo {pages} {205437} (\bibinfo {year}
  {2016})}\BibitemShut {NoStop}%
\bibitem [{\citenamefont {Fuchs}\ \emph {et~al.}(2018)\citenamefont {Fuchs},
  \citenamefont {Mosseri},\ and\ \citenamefont {Vidal}}]{TopoQC2}%
  \BibitemOpen
  \bibfield  {author} {\bibinfo {author} {\bibfnamefont {J.-N.}\ \bibnamefont
  {Fuchs}}, \bibinfo {author} {\bibfnamefont {R.}~\bibnamefont {Mosseri}},\
  and\ \bibinfo {author} {\bibfnamefont {J.}~\bibnamefont {Vidal}},\ }\bibfield
   {title} {\bibinfo {title} {Landau levels in quasicrystals},\ }\href
  {https://doi.org/10.1103/PhysRevB.98.165427} {\bibfield  {journal} {\bibinfo
  {journal} {Phys. Rev. B}\ }\textbf {\bibinfo {volume} {98}},\ \bibinfo
  {pages} {165427} (\bibinfo {year} {2018})}\BibitemShut {NoStop}%
\bibitem [{\citenamefont {Fulga}\ \emph {et~al.}(2016)\citenamefont {Fulga},
  \citenamefont {Pikulin},\ and\ \citenamefont {Loring}}]{TopoQC3}%
  \BibitemOpen
  \bibfield  {author} {\bibinfo {author} {\bibfnamefont {I.~C.}\ \bibnamefont
  {Fulga}}, \bibinfo {author} {\bibfnamefont {D.~I.}\ \bibnamefont {Pikulin}},\
  and\ \bibinfo {author} {\bibfnamefont {T.~A.}\ \bibnamefont {Loring}},\
  }\bibfield  {title} {\bibinfo {title} {Aperiodic weak topological
  superconductors},\ }\href {https://doi.org/10.1103/PhysRevLett.116.257002}
  {\bibfield  {journal} {\bibinfo  {journal} {Phys. Rev. Lett.}\ }\textbf
  {\bibinfo {volume} {116}},\ \bibinfo {pages} {257002} (\bibinfo {year}
  {2016})}\BibitemShut {NoStop}%
\bibitem [{\citenamefont {Huang}\ and\ \citenamefont
  {Liu}(2018{\natexlab{a}})}]{TopoQC4}%
  \BibitemOpen
  \bibfield  {author} {\bibinfo {author} {\bibfnamefont {H.}~\bibnamefont
  {Huang}}\ and\ \bibinfo {author} {\bibfnamefont {F.}~\bibnamefont {Liu}},\
  }\bibfield  {title} {\bibinfo {title} {Quantum spin hall effect and spin bott
  index in a quasicrystal lattice},\ }\href
  {https://doi.org/10.1103/PhysRevLett.121.126401} {\bibfield  {journal}
  {\bibinfo  {journal} {Phys. Rev. Lett.}\ }\textbf {\bibinfo {volume} {121}},\
  \bibinfo {pages} {126401} (\bibinfo {year} {2018}{\natexlab{a}})}\BibitemShut
  {NoStop}%
\bibitem [{\citenamefont {Huang}\ and\ \citenamefont
  {Liu}(2018{\natexlab{b}})}]{TopoQC5}%
  \BibitemOpen
  \bibfield  {author} {\bibinfo {author} {\bibfnamefont {H.}~\bibnamefont
  {Huang}}\ and\ \bibinfo {author} {\bibfnamefont {F.}~\bibnamefont {Liu}},\
  }\bibfield  {title} {\bibinfo {title} {Theory of spin bott index for quantum
  spin hall states in nonperiodic systems},\ }\href
  {https://doi.org/10.1103/PhysRevB.98.125130} {\bibfield  {journal} {\bibinfo
  {journal} {Phys. Rev. B}\ }\textbf {\bibinfo {volume} {98}},\ \bibinfo
  {pages} {125130} (\bibinfo {year} {2018}{\natexlab{b}})}\BibitemShut
  {NoStop}%
\bibitem [{\citenamefont {Varjas}\ \emph {et~al.}(2019)\citenamefont {Varjas},
  \citenamefont {Lau}, \citenamefont {Poyhonen}, \citenamefont {Akhmerov},
  \citenamefont {Pikulin},\ and\ \citenamefont {Fulga}}]{TopoQC6}%
  \BibitemOpen
  \bibfield  {author} {\bibinfo {author} {\bibfnamefont {D.}~\bibnamefont
  {Varjas}}, \bibinfo {author} {\bibfnamefont {A.}~\bibnamefont {Lau}},
  \bibinfo {author} {\bibfnamefont {K.}~\bibnamefont {Poyhonen}}, \bibinfo
  {author} {\bibfnamefont {A.~R.}\ \bibnamefont {Akhmerov}}, \bibinfo {author}
  {\bibfnamefont {D.~I.}\ \bibnamefont {Pikulin}},\ and\ \bibinfo {author}
  {\bibfnamefont {I.~C.}\ \bibnamefont {Fulga}},\ }\bibfield  {title} {\bibinfo
  {title} {Topological phases without crystalline counterparts},\ }\href
  {https://arxiv.org/abs/1904.07242} {\bibfield  {journal} {\bibinfo  {journal}
  {arXiv:1904.07242}\ } (\bibinfo {year} {2019})}\BibitemShut {NoStop}%
\bibitem [{\citenamefont {Chen}\ \emph
  {et~al.}(2019{\natexlab{a}})\citenamefont {Chen}, \citenamefont {Chen},
  \citenamefont {Gao}, \citenamefont {Zhou},\ and\ \citenamefont
  {Xu}}]{TopoQC61}%
  \BibitemOpen
  \bibfield  {author} {\bibinfo {author} {\bibfnamefont {R.}~\bibnamefont
  {Chen}}, \bibinfo {author} {\bibfnamefont {C.~Z.}\ \bibnamefont {Chen}},
  \bibinfo {author} {\bibfnamefont {J.~H.}\ \bibnamefont {Gao}}, \bibinfo
  {author} {\bibfnamefont {B.}~\bibnamefont {Zhou}},\ and\ \bibinfo {author}
  {\bibfnamefont {D.~H.}\ \bibnamefont {Xu}},\ }\bibfield  {title} {\bibinfo
  {title} {Higher-order topological insulators in quasicrystals},\ }\href
  {https://arxiv.org/abs/1904.09932} {\bibfield  {journal} {\bibinfo  {journal}
  {arXiv:1904.09932}\ } (\bibinfo {year} {2019}{\natexlab{a}})}\BibitemShut
  {NoStop}%
\bibitem [{\citenamefont {Bandres}\ \emph {et~al.}(2016)\citenamefont
  {Bandres}, \citenamefont {Rechtsman},\ and\ \citenamefont {Segev}}]{TopoQC7}%
  \BibitemOpen
  \bibfield  {author} {\bibinfo {author} {\bibfnamefont {M.~A.}\ \bibnamefont
  {Bandres}}, \bibinfo {author} {\bibfnamefont {M.~C.}\ \bibnamefont
  {Rechtsman}},\ and\ \bibinfo {author} {\bibfnamefont {M.}~\bibnamefont
  {Segev}},\ }\bibfield  {title} {\bibinfo {title} {Topological photonic
  quasicrystals: Fractal topological spectrum and protected transport},\ }\href
  {https://doi.org/10.1103/PhysRevX.6.011016} {\bibfield  {journal} {\bibinfo
  {journal} {Phys. Rev. X}\ }\textbf {\bibinfo {volume} {6}},\ \bibinfo {pages}
  {011016} (\bibinfo {year} {2016})}\BibitemShut {NoStop}%
\bibitem [{\citenamefont {Kitaev}(2006)}]{Kitaev}%
  \BibitemOpen
  \bibfield  {author} {\bibinfo {author} {\bibfnamefont {A.}~\bibnamefont
  {Kitaev}},\ }\bibfield  {title} {\bibinfo {title} {Anyons in an exactly
  solved model and beyond},\ }\href
  {https://doi.org/https://doi.org/10.1016/j.aop.2005.10.005} {\bibfield
  {journal} {\bibinfo  {journal} {Annals of Physics}\ }\textbf {\bibinfo
  {volume} {321}},\ \bibinfo {pages} {2 } (\bibinfo {year} {2006})}\BibitemShut
  {NoStop}%
\bibitem [{\citenamefont {Loring}\ and\ \citenamefont
  {Hastings}(2010)}]{Calgebra}%
  \BibitemOpen
  \bibfield  {author} {\bibinfo {author} {\bibfnamefont {T.~A.}\ \bibnamefont
  {Loring}}\ and\ \bibinfo {author} {\bibfnamefont {M.~B.}\ \bibnamefont
  {Hastings}},\ }\bibfield  {title} {\bibinfo {title} {Disordered topological
  insulators via c*-algebras},\ }\href
  {https://doi.org/10.1209/0295-5075/92/67004} {\bibfield  {journal} {\bibinfo
  {journal} {{EPL} (Europhysics Letters)}\ }\textbf {\bibinfo {volume} {92}},\
  \bibinfo {pages} {67004} (\bibinfo {year} {2010})}\BibitemShut {NoStop}%
\bibitem [{\citenamefont {Bianco}\ and\ \citenamefont {Resta}(2011)}]{LCHN1}%
  \BibitemOpen
  \bibfield  {author} {\bibinfo {author} {\bibfnamefont {R.}~\bibnamefont
  {Bianco}}\ and\ \bibinfo {author} {\bibfnamefont {R.}~\bibnamefont {Resta}},\
  }\bibfield  {title} {\bibinfo {title} {Mapping topological order in
  coordinate space},\ }\href {https://doi.org/10.1103/PhysRevB.84.241106}
  {\bibfield  {journal} {\bibinfo  {journal} {Phys. Rev. B}\ }\textbf {\bibinfo
  {volume} {84}},\ \bibinfo {pages} {241106} (\bibinfo {year}
  {2011})}\BibitemShut {NoStop}%
\bibitem [{\citenamefont {Brzezi\ifmmode~\acute{n}\else \'{n}\fi{}ska}\ \emph
  {et~al.}(2018)\citenamefont {Brzezi\ifmmode~\acute{n}\else \'{n}\fi{}ska},
  \citenamefont {Cook},\ and\ \citenamefont {Neupert}}]{Sierp}%
  \BibitemOpen
  \bibfield  {author} {\bibinfo {author} {\bibfnamefont {M.}~\bibnamefont
  {Brzezi\ifmmode~\acute{n}\else \'{n}\fi{}ska}}, \bibinfo {author}
  {\bibfnamefont {A.~M.}\ \bibnamefont {Cook}},\ and\ \bibinfo {author}
  {\bibfnamefont {T.}~\bibnamefont {Neupert}},\ }\bibfield  {title} {\bibinfo
  {title} {Topology in the sierpi\ifmmode \acute{n}\else
  \'{n}\fi{}ski-hofstadter problem},\ }\href
  {https://doi.org/10.1103/PhysRevB.98.205116} {\bibfield  {journal} {\bibinfo
  {journal} {Phys. Rev. B}\ }\textbf {\bibinfo {volume} {98}},\ \bibinfo
  {pages} {205116} (\bibinfo {year} {2018})}\BibitemShut {NoStop}%
\bibitem [{\citenamefont {Hasan}\ and\ \citenamefont {Kane}(2010)}]{Kane}%
  \BibitemOpen
  \bibfield  {author} {\bibinfo {author} {\bibfnamefont {M.~Z.}\ \bibnamefont
  {Hasan}}\ and\ \bibinfo {author} {\bibfnamefont {C.~L.}\ \bibnamefont
  {Kane}},\ }\bibfield  {title} {\bibinfo {title} {Colloquium: Topological
  insulators},\ }\href {https://doi.org/10.1103/RevModPhys.82.3045} {\bibfield
  {journal} {\bibinfo  {journal} {Rev. Mod. Phys.}\ }\textbf {\bibinfo {volume}
  {82}},\ \bibinfo {pages} {3045} (\bibinfo {year} {2010})}\BibitemShut
  {NoStop}%
\bibitem [{\citenamefont {Qi}\ and\ \citenamefont {Zhang}(2011)}]{Qi}%
  \BibitemOpen
  \bibfield  {author} {\bibinfo {author} {\bibfnamefont {X.-L.}\ \bibnamefont
  {Qi}}\ and\ \bibinfo {author} {\bibfnamefont {S.-C.}\ \bibnamefont {Zhang}},\
  }\bibfield  {title} {\bibinfo {title} {Topological insulators and
  superconductors},\ }\href {https://doi.org/10.1103/RevModPhys.83.1057}
  {\bibfield  {journal} {\bibinfo  {journal} {Rev. Mod. Phys.}\ }\textbf
  {\bibinfo {volume} {83}},\ \bibinfo {pages} {1057} (\bibinfo {year}
  {2011})}\BibitemShut {NoStop}%
\bibitem [{\citenamefont {Groth}\ \emph {et~al.}(2014)\citenamefont {Groth},
  \citenamefont {Wimmer}, \citenamefont {Akhmerov},\ and\ \citenamefont
  {Waintal}}]{Kwant}%
  \BibitemOpen
  \bibfield  {author} {\bibinfo {author} {\bibfnamefont {C.~W.}\ \bibnamefont
  {Groth}}, \bibinfo {author} {\bibfnamefont {M.}~\bibnamefont {Wimmer}},
  \bibinfo {author} {\bibfnamefont {A.~R.}\ \bibnamefont {Akhmerov}},\ and\
  \bibinfo {author} {\bibfnamefont {X.}~\bibnamefont {Waintal}},\ }\bibfield
  {title} {\bibinfo {title} {Kwant: a software package for quantum transport},\
  }\href {https://doi.org/10.1088/1367-2630/16/6/063065} {\bibfield  {journal}
  {\bibinfo  {journal} {New Journal of Physics}\ }\textbf {\bibinfo {volume}
  {16}},\ \bibinfo {pages} {063065} (\bibinfo {year} {2014})}\BibitemShut
  {NoStop}%
\bibitem [{\citenamefont {{Landauer}}(1957)}]{Landauer}%
  \BibitemOpen
  \bibfield  {author} {\bibinfo {author} {\bibfnamefont {R.}~\bibnamefont
  {{Landauer}}},\ }\bibfield  {title} {\bibinfo {title} {Spatial variation of
  currents and fields due to localized scatterers in metallic conduction},\
  }\href {https://doi.org/10.1147/rd.13.0223} {\bibfield  {journal} {\bibinfo
  {journal} {IBM Journal of Research and Development}\ }\textbf {\bibinfo
  {volume} {1}},\ \bibinfo {pages} {223} (\bibinfo {year} {1957})}\BibitemShut
  {NoStop}%
\bibitem [{\citenamefont {B\"uttiker}(1988)}]{Buttiker}%
  \BibitemOpen
  \bibfield  {author} {\bibinfo {author} {\bibfnamefont {M.}~\bibnamefont
  {B\"uttiker}},\ }\bibfield  {title} {\bibinfo {title} {Absence of
  backscattering in the quantum hall effect in multiprobe conductors},\ }\href
  {https://doi.org/10.1103/PhysRevB.38.9375} {\bibfield  {journal} {\bibinfo
  {journal} {Phys. Rev. B}\ }\textbf {\bibinfo {volume} {38}},\ \bibinfo
  {pages} {9375} (\bibinfo {year} {1988})}\BibitemShut {NoStop}%
\bibitem [{\citenamefont {Datta}(2005)}]{Datta}%
  \BibitemOpen
  \bibfield  {author} {\bibinfo {author} {\bibfnamefont {S.}~\bibnamefont
  {Datta}},\ }\href@noop {} {\emph {\bibinfo {title} {Quantum Transport: Atom
  to Transistor}}}\ (\bibinfo  {publisher} {Cambridge University Press},\
  \bibinfo {year} {2005})\BibitemShut {NoStop}%
\bibitem [{\citenamefont {Chen}\ \emph
  {et~al.}(2019{\natexlab{b}})\citenamefont {Chen}, \citenamefont {Xu},\ and\
  \citenamefont {Zhou}}]{transport}%
  \BibitemOpen
  \bibfield  {author} {\bibinfo {author} {\bibfnamefont {R.}~\bibnamefont
  {Chen}}, \bibinfo {author} {\bibfnamefont {D.-H.}\ \bibnamefont {Xu}},\ and\
  \bibinfo {author} {\bibfnamefont {B.}~\bibnamefont {Zhou}},\ }\bibfield
  {title} {\bibinfo {title} {Topological anderson insulator phase in a
  quasicrystal lattice},\ }\href {https://doi.org/10.1103/PhysRevB.100.115311}
  {\bibfield  {journal} {\bibinfo  {journal} {Phys. Rev. B}\ }\textbf {\bibinfo
  {volume} {100}},\ \bibinfo {pages} {115311} (\bibinfo {year}
  {2019}{\natexlab{b}})}\BibitemShut {NoStop}%
\bibitem [{\citenamefont {Dang}\ \emph {et~al.}(2016)\citenamefont {Dang},
  \citenamefont {Burton},\ and\ \citenamefont {Tsymbal}}]{Faliue_HC1}%
  \BibitemOpen
  \bibfield  {author} {\bibinfo {author} {\bibfnamefont {X.}~\bibnamefont
  {Dang}}, \bibinfo {author} {\bibfnamefont {J.~D.}\ \bibnamefont {Burton}},\
  and\ \bibinfo {author} {\bibfnamefont {E.~Y.}\ \bibnamefont {Tsymbal}},\
  }\bibfield  {title} {\bibinfo {title} {Magnetic gating of a 2d topological
  insulator},\ }\href {https://doi.org/10.1088/0953-8984/28/38/38lt01}
  {\bibfield  {journal} {\bibinfo  {journal} {Journal of Physics: Condensed
  Matter}\ }\textbf {\bibinfo {volume} {28}},\ \bibinfo {pages} {38LT01}
  (\bibinfo {year} {2016})}\BibitemShut {NoStop}%
\bibitem [{\citenamefont {Zheng}\ and\ \citenamefont
  {Cazalilla}(2018)}]{Faliue_HC2}%
  \BibitemOpen
  \bibfield  {author} {\bibinfo {author} {\bibfnamefont {J.-H.}\ \bibnamefont
  {Zheng}}\ and\ \bibinfo {author} {\bibfnamefont {M.~A.}\ \bibnamefont
  {Cazalilla}},\ }\bibfield  {title} {\bibinfo {title} {Nontrivial interplay of
  strong disorder and interactions in quantum spin-hall insulators doped with
  dilute magnetic impurities},\ }\href
  {https://doi.org/10.1103/PhysRevB.97.235402} {\bibfield  {journal} {\bibinfo
  {journal} {Phys. Rev. B}\ }\textbf {\bibinfo {volume} {97}},\ \bibinfo
  {pages} {235402} (\bibinfo {year} {2018})}\BibitemShut {NoStop}%
\bibitem [{\citenamefont {Novelli}\ \emph {et~al.}(2019)\citenamefont
  {Novelli}, \citenamefont {Taddei}, \citenamefont {Geim},\ and\ \citenamefont
  {Polini}}]{Faliue_HC3}%
  \BibitemOpen
  \bibfield  {author} {\bibinfo {author} {\bibfnamefont {P.}~\bibnamefont
  {Novelli}}, \bibinfo {author} {\bibfnamefont {F.}~\bibnamefont {Taddei}},
  \bibinfo {author} {\bibfnamefont {A.~K.}\ \bibnamefont {Geim}},\ and\
  \bibinfo {author} {\bibfnamefont {M.}~\bibnamefont {Polini}},\ }\bibfield
  {title} {\bibinfo {title} {Failure of conductance quantization in
  two-dimensional topological insulators due to nonmagnetic impurities},\
  }\href {https://doi.org/10.1103/PhysRevLett.122.016601} {\bibfield  {journal}
  {\bibinfo  {journal} {Phys. Rev. Lett.}\ }\textbf {\bibinfo {volume} {122}},\
  \bibinfo {pages} {016601} (\bibinfo {year} {2019})}\BibitemShut {NoStop}%
\bibitem [{\citenamefont {Verbin}\ \emph {et~al.}(2013)\citenamefont {Verbin},
  \citenamefont {Zilberberg}, \citenamefont {Kraus}, \citenamefont {Lahini},\
  and\ \citenamefont {Silberberg}}]{TP_OPQC}%
  \BibitemOpen
  \bibfield  {author} {\bibinfo {author} {\bibfnamefont {M.}~\bibnamefont
  {Verbin}}, \bibinfo {author} {\bibfnamefont {O.}~\bibnamefont {Zilberberg}},
  \bibinfo {author} {\bibfnamefont {Y.~E.}\ \bibnamefont {Kraus}}, \bibinfo
  {author} {\bibfnamefont {Y.}~\bibnamefont {Lahini}},\ and\ \bibinfo {author}
  {\bibfnamefont {Y.}~\bibnamefont {Silberberg}},\ }\bibfield  {title}
  {\bibinfo {title} {Observation of topological phase transitions in photonic
  quasicrystals},\ }\href {https://doi.org/10.1103/PhysRevLett.110.076403}
  {\bibfield  {journal} {\bibinfo  {journal} {Phys. Rev. Lett.}\ }\textbf
  {\bibinfo {volume} {110}},\ \bibinfo {pages} {076403} (\bibinfo {year}
  {2013})}\BibitemShut {NoStop}%
\bibitem [{\citenamefont {Ahn}\ \emph {et~al.}(2018)\citenamefont {Ahn},
  \citenamefont {Moon}, \citenamefont {Kim}, \citenamefont {Kim}, \citenamefont
  {Shin}, \citenamefont {Kim}, \citenamefont {Cha}, \citenamefont {Kahng},
  \citenamefont {Kim}, \citenamefont {Koshino}, \citenamefont {Son},
  \citenamefont {Yang},\ and\ \citenamefont {Ahn}}]{Ahn782}%
  \BibitemOpen
  \bibfield  {author} {\bibinfo {author} {\bibfnamefont {S.~J.}\ \bibnamefont
  {Ahn}}, \bibinfo {author} {\bibfnamefont {P.}~\bibnamefont {Moon}}, \bibinfo
  {author} {\bibfnamefont {T.-H.}\ \bibnamefont {Kim}}, \bibinfo {author}
  {\bibfnamefont {H.-W.}\ \bibnamefont {Kim}}, \bibinfo {author} {\bibfnamefont
  {H.-C.}\ \bibnamefont {Shin}}, \bibinfo {author} {\bibfnamefont {E.~H.}\
  \bibnamefont {Kim}}, \bibinfo {author} {\bibfnamefont {H.~W.}\ \bibnamefont
  {Cha}}, \bibinfo {author} {\bibfnamefont {S.-J.}\ \bibnamefont {Kahng}},
  \bibinfo {author} {\bibfnamefont {P.}~\bibnamefont {Kim}}, \bibinfo {author}
  {\bibfnamefont {M.}~\bibnamefont {Koshino}}, \bibinfo {author} {\bibfnamefont
  {Y.-W.}\ \bibnamefont {Son}}, \bibinfo {author} {\bibfnamefont {C.-W.}\
  \bibnamefont {Yang}},\ and\ \bibinfo {author} {\bibfnamefont {J.~R.}\
  \bibnamefont {Ahn}},\ }\bibfield  {title} {\bibinfo {title} {Dirac electrons
  in a dodecagonal graphene quasicrystal},\ }\href
  {https://doi.org/10.1126/science.aar8412} {\bibfield  {journal} {\bibinfo
  {journal} {Science}\ }\textbf {\bibinfo {volume} {361}},\ \bibinfo {pages}
  {782} (\bibinfo {year} {2018})}\BibitemShut {NoStop}%
\end{thebibliography}%

\end{document}